\documentclass[12pt]{article}

\pdfoutput=1

\usepackage[usenames,dvips]{color}
\usepackage{amssymb}
\usepackage{amsmath}
\usepackage{epsfig}
\usepackage[utf8]{inputenc}
\usepackage[T1]{fontenc}
\usepackage{graphicx}
\usepackage{xcolor}

\author{P.~Kozów and M.~Olechowski\\[12pt]
\em
Institute of Theoretical Physics, Faculty of Physics, University of Warsaw\\[6pt]
\em Pasteura 5, 02-093 Warsaw, Poland
}
\date{}
\title{\bf Early universe dynamics of PQ field\\ 
with very small self-coupling\\ 
and its implications for axion dark matter}

\begin{document}
\thispagestyle{empty}

\maketitle
\begin{abstract}
Axion-like particles (ALPs) are often considered as good candidates for dark matter. Several mechanisms generating relic abundance of ALP dark matter have been proposed. They may involve processes which take place before, during or after cosmic inflation. In all cases an important role is played by the potential of the corresponding Peccei-Quinn (PQ) field. Quite often this potential is assumed to be dominated by a quartic term with a very small coupling. We show that in such situation it is crucial to take into account different kinds of corrections especially in models in which the PQ field evolves during and after inflation. We investigate how such evolution changes due to radiative, thermal and geometric corrections. In many cases those changes are very important and result in strong modifications of the predictions of a model. They may strongly influence the amount of ALP contributions to cold and warm components of dark matter as well as the power spectrum of associated isocurvature perturbations. Models with a quasi-supersymmetric spectrum of particles to which the PQ field couples seem to be especially interesting. 
Qualitative features of such models are discussed with the help of approximate analytical formulae. However, the dynamics of the PQ field with the considered corrections taken into account is more complicated than in the case without corrections so dedicated numerical calculations are necessary to obtain precise predictions.
We present such results for some characteristic benchmark points in the parameter space.
\end{abstract}


\section{Introduction}
\label{sec:intro}

In this work we consider general axion-like particles (ALPs) to which we will refer as axions. By QCD axion we denote such axion which is used to solve the problem of CP symmetry breaking in strong interactions. An axion field is related to some complex Peccei-Quinn (PQ) scalar field, $\Phi=\frac{1}{\sqrt{2}}S e^{i\theta}$, charged under a global ${U(1)_{PQ}}$ symmetry \cite{Peccei:1977hh,Peccei:1977ur}. The phase of $\Phi$ may be expressed as $\theta=a/f_a$ where $a$ is the axion field and $f_a$ is the axion decay constant. We will use the name saxion for the radial component $S$. ${U(1)_{PQ}}$ is anomalous which leads to generation of an axion potential via non-perturbative effects. Axions are considered as very interesting candidates for dark matter (DM) in the universe 
(see e.g.~reviews \cite{Raffelt:1990yz,Visinelli:2009zm,Ringwald:2012hr,Marsh:2015xka,DiLuzio:2020wdo,Irastorza:2021tdu} and references therein). In most of the proposed models they play the role of cold dark matter (CDM) but there are also proposals \cite{Co:2017mop,Co:2020dya} in which axions behave as warm dark matter (WDM). In this work we are mostly interested in models in which one axion contributes to both CDM and WDM. Phenomenological aspects of models with a mixture of some CDM and WDM were quite intensively studied \cite{Kamada:2016vsc,Diamanti:2017xfo,Gariazzo:2017pzb,Parimbelli:2021mtp,Marsh:2013ywa}. Several models were proposed in which axion is one component of mixed DM but usually with WIMP  \cite{Acharya:2012dz,Baer:2011hx,Bae:2015rra} or axino \cite{Baer:2008yd,Bae:2011jb} being the second component.

A scenario in which axion contributes to both CDM and WDM may be realized when the corresponding PQ field undergoes some non-trivial evolution during and/or after inflation. 
Usually one assumes that the potential (or at least its leading part) for the PQ field $\Phi$ has the simplest form leading to spontaneous breaking of the ${U(1)_{PQ}}$ symmetry, namely
\begin{equation}
V(\Phi)=\lambda_{\Phi}\left(\left|\Phi\right|^2-\frac{f_a^2}{2}\right)^2=\frac{\lambda_{\Phi}}{4}\left(S^2-f_a^2\right)^2\,.
\label{Vlambda4}
\end{equation}
The position of the minimum of this potential is at $S_{\rm min}=f_a$ and the mass squared of the radial mode at this minimum equals $m_S^2=2\lambda_\Phi f_a^2$.

Models with non-trivial dynamics of $\Phi$ during inflation have been proposed and investigated \cite{Moroi:2013tea,Graham:2018jyp,Takahashi:2018tdu, Co:2019wyp,Co:2019jts,Moroi:2020has,Co:2021lkc,Gouttenoire:2021jhk}. Typically in such models the self-coupling $\lambda_\Phi$ must be very small in order to avoid too big isocurvature perturbations related to the relic abundance of axions. Values of $\lambda_\Phi$ considered in the literature are smaller
(sometimes even by many orders of magnitude) than $10^{-20}$. In such situation one should consider corrections to the potential \eqref{Vlambda4} and check how they may modify predictions of a given model. One type of such corrections, namely non-renormalizable corrections breaking ${U(1)_{PQ}}$, has been investigated in 
\cite{Co:2020dya,Co:2019wyp,Co:2019jts} and used as a crucial ingredient of the kinematic misalignment mechanism. In the present work we investigate possible consequences of other types of corrections: radiative, geometric (related to the curvature of space-time) and thermal.

It occurs that such corrections strongly modify the dynamics of the PQ field. Due to geometrical corrections the evolution of the saxion component during inflation may have character quite different from that for the simple potential \eqref{Vlambda4}. Interplay between geometric and thermal corrections leads to a quite rich spectrum of scenarios which may be realized after inflation. Different scenarios, obtained for different sets of parameters of the model, lead to different axion contributions to CDM and WDM. In many cases the amount of axion WDM 
may change even by orders of magnitude when the corrections are taken into account. Also the power spectrum of associated isocurvature perturbations may be quite different.
However, due to the complexity of the system, dedicated numerical simulations are necessary to obtain precise quantitative results.


\section{Peccei-Quinn symmetry broken by Coleman-Weinberg mechanism}

Let us start with the potential for the Peccei-Quinn field at late stages of the universe evolution when one may neglect effects caused by non-zero temperature and non-zero curvature of the space-time. In such situation we have to take into account only the usual radiative corrections to the potential \eqref{Vlambda4}. These corrections may be written in the form of the  Coleman-Weinberg (CW) potential \cite{Coleman:1973jx}. The coupling $\lambda_\Phi$ is very small so it is natural to use the approach proposed by Gildner and Weinberg \cite{Gildener:1976ih}, i.e.~to use the renormalization scale $\mu$ at which this self-coupling constant vanishes  $\lambda_\Phi(\mu)=0$. Contributions to the CW potential for the PQ field come from each field to which $\Phi$ couples. $\Phi$ must couple to some fermions in order to make ${U(1)_{PQ}}$ anomalous. However, fermion contributions tend to destabilize the CW potential for large values of $\Phi$. Thus, we consider models in which the PQ scalar couples also to some bosons\footnote{Another alternative could be to assume some non-renormalizable terms of higher order in $\Phi$.}
which for simplicity we choose to be scalars. These couplings are described by the following terms in the Lagrangian
\begin{equation}
{\cal{L}}\supset -\sum_i \frac12\lambda_{\Phi\phi_i}\left|\Phi\right|^2\phi_i^2-\sum_j y_j\Phi\overline{\psi}_j\psi_j
\label{L}
\end{equation}
and lead to the CW potential:
\begin{equation}
V(\Phi)=\frac{1}{64\pi^2} \sum_{\rm scalars}  M_{\phi_i}^4\left[\ln\left(\frac{M_{\phi_i}^2}{\mu^2}\right)-\frac32\right]
-\frac{4}{64\pi^2} \sum_{\rm fermions}  M_{\psi_j}^4\left[\ln\left(\frac{M_{\psi_j}^2}{\mu^2}\right)-\frac32\right].
\label{VCW}
\end{equation}
The SM Higgs field may be among scalars $\phi_i$ coupled to $\Phi$. 
The scalar and fermion masses in the above formula depend on $\Phi$. From \eqref{L} and the scalar mass terms $\frac12m_i^2\phi_i^2$ we get
\begin{equation}
M_{\phi_i}^2=m_i^2+\lambda_{\Phi\phi_i}\left|\Phi\right|^2
\,,\qquad
M_{\psi_j}^2=y_j^2\left|\Phi\right|^2\,.
\end{equation}

The CW potential \eqref{VCW} is bounded from below at large values of $|\Phi|$ if the contribution from scalars dominate over that from fermions. 
It occurs that the model has several interesting features when the scalar contribution dominates only slightly. The most interesting situation is obtained when the spectrum of particles to which the PQ field couples is similar to a supersymmetric one\footnote{It may follow for example from a hidden sector with supersymmetry softly broken at some high scale. Different running of bosonic and fermionic couplings may result in a situation when these couplings are similar but not exactly equal.}. 
For simplicity we assume there are $N_f$ fermions and $N_s=4N_f$ scalars with the coupling constants and masses satisfying the conditions: $y_j=y$, $\lambda_{\Phi\phi_i}=\lambda$, $m_i^2=m^2$. 
Quasi-supersymmetric nature of the spectrum means that the fermion and scalar couplings are not very different: $\lambda\approx y^2$. Let us introduce parameter $\delta$ such that
\begin{equation}
y^2=(1-\delta)\lambda\,,
\end{equation}
where $\delta$ is smaller than 1 and not very close to 1. 
We consider only positive values of $\delta$ because otherwise the fermionic contribution in \eqref{VCW} would destabilize the potential at large values of $|\Phi|$.
The potential \eqref{VCW} takes the form
\begin{equation}
V_{0}
=
\frac{N_s}{64\pi^2}
\left[
\left(m^2+\lambda|\Phi|^2\right)^2\left(\ln\left(\frac{m^2+\lambda|\Phi|^2}{\mu^2}\right)-\frac32\right)
-(1-\delta)^2\lambda^2|\Phi|^4\left(\ln\left(\frac{(1-\delta)\lambda|\Phi|^2}{\mu^2}\right)-\frac32\right)
\right].
\label{VCW_SUSY}
\end{equation}
This potential has a minimum for $\Phi\ne0$ only if $m^2<e\mu^2$. 
For small $\delta$ the position of such minimum is approximately equal to
\begin{equation}
S_{\rm min, 0}^2 \approx
2\,\frac{\mu^2}{\lambda}\left[1-\frac{m^2}{2\mu^2}+\frac{m^4}{24\mu^4}
+\delta\left(\frac{\mu^2}{m^2}+\frac12-\frac{m^2}{3\mu^2}\right)\right],
\label{Smin^2_SUSY_0}
\end{equation}
while its depth may be well approximated by 
\begin{equation}
\Delta V_0 \approx 
\frac{N_s\mu^4}{32\pi^2}\left\{
\frac{m^2}{\mu^2}\left[1+\frac{m^2}{2\mu^2}\left(\ln\left(\frac{m^2}{\mu^2}\right)-\frac32\right)
-\frac{m^4}{24\mu^2}
\right]
+\delta\left(1-\frac{m^2}{2\mu^2}
-\frac{m^4}{12\mu^4}
\right)\right\}\,.
\label{DeltaV_SUSY_0}
\end{equation}
The saxion mass reads
\begin{equation}
m_S^2
\approx
\frac{N_s\lambda^2}{32\pi^2}\frac{m^2}{\mu^2}S_{\rm min,0}^2\,.
\label{mS^2_SUSY_0}
\end{equation}

Let us compare the above results to those for the usually considered potential \eqref{Vlambda4}, for which $S_{\rm min}^2=f_a^2$ and $m_S^2/S_{\rm min}^2=2\lambda_\Phi$. One can see that the position of the minimum 
corresponds to the replacement of the axion decay constant $f_a^2$ by an expression of order ${\mu^2}/{\lambda}$. 
The ratio of the saxion mass to the axion decay constant, $m_S^2/S_{\rm min}^2$, may be approximately reproduced by the replacement $\lambda_\Phi\to\frac{N_s\lambda^2}{64\pi^2}\frac{m^2}{\mu^2}$.

Let us now take into account thermal corrections to the potential while still neglecting the curvature of space-time. We would like to estimate the critical temperature at which the PQ symmetry may be restored. It may be defined as the minimal temperature $T_{c}$ for which the global minimum of the potential is located at $\Phi=0$.

The thermal contribution to the mass term of the PQ field may be written as
\begin{equation}
\frac{\alpha}{24}T^2 |\Phi|^2\,,
\label{thermal_mass}
\end{equation}
where in our model
\begin{equation}
\alpha = \sum_{\rm scalars} \lambda_{\Phi\phi_i}+2\sum_{\rm fermions}y_j^2\,,
\label{alpha}
\end{equation}
with sums over bosons and fermions which are in thermal equilibrium and have masses not (much) bigger than the temperature $T$. Particles substantially heavier than $T$ and particles which are not in perfect equilibrium may also partially contribute to \eqref{thermal_mass}. We describe all possible situations by introducing an effective number of degrees of freedom, $n_{\rm eff}$, contributing to the PQ field mass term 
\begin{equation}
\alpha=n_{\rm eff}\lambda\,.
\label{neff}
\end{equation}
Maximal value of $n_{\rm eff}$ is close to 
$\frac32N_s$ 
when all particles, $\phi_i$ and $\psi_j$, are in thermal equilibrium and are light enough. In the limit of all these particles completely decoupled $n_{\rm eff}$ tends to zero.

One should remember that $n_{\rm eff}$ is not just a constant. The effective number of degrees of freedom of particles $\phi$ and $\psi$ which are not totally decoupled from thermal bath depends on masses and couplings of those particles and on temperature. We use it as a convenient tool do discuss the leading effects caused of thermal corrections. The full thermal potential should be used when more precise quantitative results are needed (see section \ref{sec:numerical_results}).

Above $T_{c}$ the temperature correction \eqref{thermal_mass} evaluated at the position of the (zero temperature) minimum \eqref{Smin^2_SUSY_0} is bigger than the depth of such minimum \eqref{DeltaV_SUSY_0}. This condition (neglecting the term proportional to $\delta$) reads
\begin{equation}
\frac{\lambda n_{\rm eff}(T)}{24}\,T^2\, \frac{\mu^2}{\lambda}\left[1-\frac{m^2}{2\mu^2}+\frac{m^4}{24\mu^4}
\right]
\gtrsim
\frac{N_s}{32\pi^2}\,
m^2\mu^2\left[1+\frac{m^2}{2\mu^2}\left(\ln\left(\frac{m^2}{\mu^2}\right)-\frac32\right)-\frac{m^4}{24\mu^2}\right].
\end{equation}
Thus, the critical temperature may be approximated by
\begin{equation}
T_{c}
\approx
m\sqrt{\frac{3N_s}{4\pi^2n_{\rm eff}(T_c)}\,
\frac{1+\frac{m^2}{2\mu^2}\left(\ln\left(\frac{m^2}{\mu^2}\right)-\frac32\right)-\frac{m^4}{24\mu^4}}{1-\frac{m^2}{2\mu^2}+\frac{m^4}{24\mu^4}}}\,.
\label{Tc}
\end{equation}
Its minimal possible value, corresponding to the maximal possible 
$n_{\rm eff}\approx\frac32N_s$, 
is of order ${\cal{O}}(0.1m)$. $T_c$ may be much higher if $n_{\rm eff}\ll1$. 
In such a case a more appropriate way to take thermal effects into account would be to use the full thermal potential or the so-called thermal logarithmic potential \cite{Anisimov:2000wx}.
However, for simplicity of our qualitative discussion, we will parameterize different thermal effects  by the effective parameter $n_{\rm eff}$.


\section{Corrections from curvature of space-time}

When the curvature of the space-time may not be neglected the expression for the CW potential \eqref{VCW} generalizes to \cite{Markkanen:2018bfx,Hardwick:2019uex}
\begin{align}
V(\Phi)=&\,\frac{1}{64\pi^2} \sum_{\rm bosons}  \left\{M_{\phi_i}^4\left[\ln\left(\frac{\left|M_{\phi_i}^2\right|}{\mu^2}\right)-\frac32\right]
+\frac{R_{\mu\nu\rho\sigma}R^{\mu\nu\rho\sigma}-R_{\mu\nu}R^{\mu\nu}}{90}\ln\left(\frac{\left|M_{\phi_i}^2\right|}{\mu^2}\right)
\right\}
\nonumber\\
&-\frac{4}{64\pi^2} \sum_{\rm fermions}  \left\{M_{\psi_j}^4\left[\ln\left(\frac{\left|M_{\psi_j}^2\right|}{\mu^2}\right)-\frac32\right]
-\frac{\frac78 R_{\mu\nu\rho\sigma}R^{\mu\nu\rho\sigma}+R_{\mu\nu}R^{\mu\nu}}{90}\ln\left(\frac{\left|M_{\psi_j}^2\right|}{\mu^2}\right)
\right\},
\label{VCW_R}
\end{align}
where $R_{\mu\nu\rho\sigma}$ and $R_{\mu\nu}$ are the Riemann and Ricci tensors, respectively. The field-dependent masses in our model are given by
\begin{equation}
M_{\phi_i}^2=m_i^2+\lambda_{\Phi\phi_i}\left|\Phi\right|^2+\left(\xi_i-\frac16\right)R
\,,\qquad
M_{\psi_j}^2=y_j^2\left|\Phi\right|^2+\frac{1}{12}R\,,
\label{MphiMpsi}
\end{equation}
where $R$ is the Ricci scalar and $\xi_i$ is the coefficient of a non-minimal coupling of the scalars $\phi_i$ to the curvature.

Neglecting the spatial curvature the curvature dependent invariants may be expressed in terms of the Hubble parameter and its time derivative:
\begin{align}
R&=6\left(\dot{H}+2H^2\right)\,,
\\
R_{\mu\nu}R^{\mu\nu}&=9\left(\dot{H}+H^2\right)^2+3\left(\dot{H}+3H^2\right)^2\,,
\\
R_{\mu\nu\rho\sigma}R^{\mu\nu\rho\sigma}&=12\left[\left(\dot{H}+H^2\right)^2+H^4\right]\,.
\end{align}
During inflation, radiation-dominated period or matter-dominated period the above formulae may be further simplified. The results are given in table \ref{table_curv-inv}.
\begin{table}[h]
\begin{center}
\begin{tabular}{|c|c|c|c|}
\hline
 & inflation & MD & RD \\
 \hline
 $R$ & $12H^2$ & $3H^2$ & $0$ \\
 \hline
 $R_{\mu\nu}R^{\mu\nu}$ & $36H^4$ & $9H^4$ & $12H^4$ \\
 \hline
 $R_{\mu\nu\rho\sigma}R^{\mu\nu\rho\sigma}$ & $24H^4$ & $15H^4$ & $24H^4$ \\
 \hline
\end{tabular}
\end{center}
 \caption{Curvature invariants as functions of the Hubble parameter during three epochs of the universe evolution: inflation, matter domination and radiation domination.}
  \label{table_curv-inv}
\end{table}

In the following we will use the potential \eqref{VCW_R} to investigate evolution of the PQ field during and after inflation when the curvature effects may play a very important role.


\subsection{CW potential during inflation}
\label{sec:G_inf}

It occurs that in many cases the curvature effects may change quite substantially the characteristics of the CW potential. Let us first discuss these effects during inflation. The potential \eqref{VCW_R} for our model with quasi-supersymmetric spectrum simplifies to 
\begin{align}
V_{\rm inf}(\Phi)
=
\frac{N_s}{64\pi^2}
&\left\{
\left[m^2+\lambda|\Phi|^2+\left(12\xi-2\right){H_I^2}\right]^2
\left[\ln\left(\frac{\left|m^2+\lambda|\Phi|^2+\left(12\xi-2\right){H_I^2}\right|}{\mu^2}\right)-\frac32\right]\right.
\nonumber\\
&-\frac{2}{15}{H_I^4}\ln\left(\frac{\left|m^2+\lambda|\Phi|^2+\left(12\xi-2\right){H_I^2}\right|}{\mu^2}\right)
\nonumber\\
&-
\left((1-\delta)\lambda|\Phi|^2+{H_I^2}\right)^2\left[\ln\left(\frac{(1-\delta)\lambda|\Phi|^2+{H_I^2}}{\mu^2}\right)-\frac32\right]
\nonumber\\
&\left.+\frac{19}{30}{H_I^4}\ln\left(\frac{(1-\delta)\lambda|\Phi|^2+{H_I^2}}{\mu^2}\right)
\right\}.
\label{VCW_inf_SUSY}
\end{align}
We have added one more simplifying assumption that all scalars $\phi_i$ have the same coupling to the Ricci scalar:
\begin{equation}
{\cal{L}}\supset -\sum_{\rm scalars} \frac12 \xi R \phi_i^2\,.
\end{equation}

\begin{figure}
\begin{center}
\includegraphics[scale=1.15]{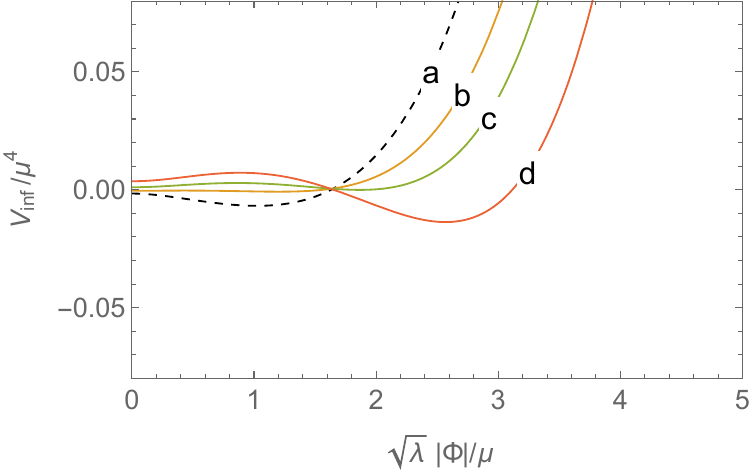}
\hfill
\includegraphics[scale=1.15]{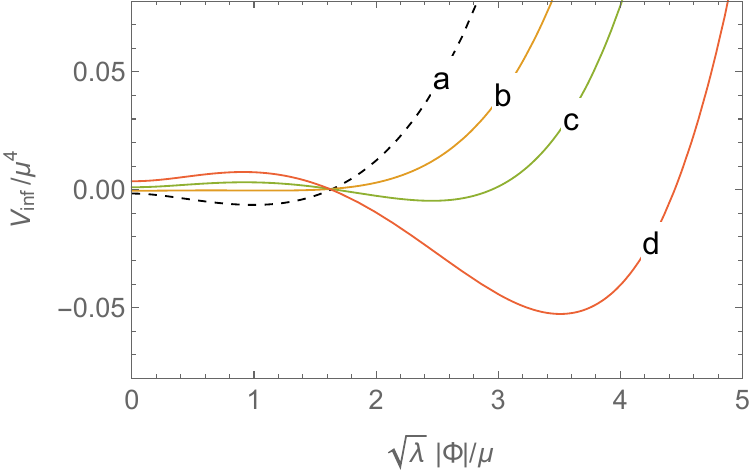}
\end{center}
\caption{Potential $V_{\rm inf}$ \eqref{VCW_inf_SUSY} as function of $\frac{\sqrt{\lambda}|\Phi|}{\mu}$ for parameters $m=0.4\mu$, $\xi=\frac16$ and for several values of ${H_I}/\mu$. Curves a, b, c and d were obtained for ${H_I}/\mu$ equal 0, 0.4, 0.5 and 0.6, respectively. The black dashed curves correspond to the CW potential \eqref{VCW_SUSY}. 
Two panels have different values of the parameter $\delta$, namely: $\delta=0.02$ on the left panel, $\delta=0.01$ on the right panel.
}
\label{F-examples}
\end{figure}

Properties of this potential depend in a quite interesting way on the value of ${H_I}$. 
For ${H_I}\to0$ we of course reproduce the late time potential \eqref{VCW_SUSY}. For small enough ${H_I}$ the $U(1)_{PQ}$ is still broken but the depth of the corresponding minimum decreases. For some range of ${H_I}$ the potential is quite flat for small $|\Phi|$. Its details change if we further increase ${H_I}$. First a local minimum develops at $|\Phi|=0$ and the PQ symmetry may be restored. For some range of ${H_I}$ it is the global minimum or even the only minimum (depending on other parameters of the model) but for bigger ${H_I}$ it is again a local one. The minimum at non-zero $|\Phi|$ becomes deeper and moves to larger values of $|\Phi|$ when ${H_I}$ is bigger. Moreover, the position and depth of such minimum depend on the parameter $\delta$. Namely, this minimum becomes deeper and moves to bigger values of $|\Phi|$ with decreasing $\delta$, i.e.~with the spectrum closer to a supersymmetric one. This behavior may be easily understood by expanding the potential \eqref{VCW_inf_SUSY} for large $|\Phi|^2$ and small $\delta$. The coefficient of the term $|\Phi|^4\ln(|\Phi|^2)$ is at least linear in $\delta$ while the coefficient of the next to the leading term, i.e.~$|\Phi|^2\ln(|\Phi|^2)$, has a part independent of $\delta$. These two most important terms give
\begin{equation}
V_{\rm inf} \approx
\frac{N_s}{32\pi^2}
\lambda|\Phi|^2\ln\left(\frac{\lambda|\Phi|^2}{e\mu^2}\right)
\left[
\delta\lambda|\Phi|^2+m^2-(3-12\xi)H_I^2
\right].
\label{Vlimit_SUSY}
\end{equation}
For $\xi<\frac14$ and big enough ${H_I}$ the $\delta$-independent term in the square bracket in the last formula is negative. This negative contribution may be compensated by the positive one with higher power of $|\Phi|^2$ which however is proportional to $\delta$. Thus, for smaller $\delta$ the leading positive term starts to dominate at larger $|\Phi|^2$. Domination of the negative term over a longer interval in $|\Phi|$ leads to a deeper minimum. Using \eqref{Vlimit_SUSY} we obtain the following approximate (in the leading order in large $|\Phi|$ and small $\delta$) expressions for the position and depth of the minimum:
\begin{align}
S^2_{\rm min,inf}
&\approx
\frac{(3-12\xi){H_I^2}-m^2}{\delta\lambda}\,,
\label{Smin^2_I}
\\
\Delta V_{\rm inf} &\approx \frac{N_s}{64\pi^2}\frac{\left[\left(3-12\xi\right){H_I^2}-m^2\right]^2}{2\delta}\,\ln\left(\frac{\left(3-12\xi\right){H_I^2}-m^2}{2e\delta\mu^2}\right)\,.
\label{DeltaV_SUSY_I}
\end{align}
All the features of the potential \eqref{VCW_inf_SUSY} discussed in this paragraph may be seen in Fig.~\ref{F-examples}. The parameters used in this figure were chosen to illustrate the qualitative features of the potential and to show some details at small values of $|\Phi|$ which allows for easy comparison with the late time CW potential \eqref{VCW_SUSY}.

The interesting features of the potential discussed above result mainly from a non-trivial interplay between the scalar and fermion parts of \eqref{VCW_inf_SUSY} in the presence of non-zero ${H_I}$. Thus, most of these features are absent 
when the curvature of space-time is small and may be neglected\footnote{For example, equations \eqref{Smin^2_SUSY_0} and \eqref{DeltaV_SUSY_0}  show that in such case the minimum of the potential for small $\delta$ only weekly depends on the exact value of $\delta$.}. Those features are absent also when the CW potential is strongly dominated by the scalar contribution (i.e.~when $1-\delta\ll1$). So, we concentrate our analysis on models with quasi-supersymmetric spectra (i.e.~when the parameter $\delta$ is not very close to 1) resulting in more interesting dynamics of the PQ field during and after inflation. Models with supersymmetry and large values of the saxion field during inflation were in a somewhat different context considered e.g.~in \cite{Hashimoto:1998ua,Kawasaki:2008jc,Co:2020jtv}.


\subsection{CW potential after inflation}

The CW potential for the PQ field \eqref{VCW_R} is almost constant during inflation. It changes only due to a slow decrease of the Hubble parameter. The situation is different after inflation because the parameters describing evolution of the universe change their values faster than during inflation. During the reheating process the evolution of the universe changes gradually from matter-dominated (energy density dominated by inflaton oscillations) to radiation-dominated (energy dominated by relativistic particles produced during reheating). The potential of the PQ field also gradually changes. Using formula \eqref{VCW_R} and table \ref{table_curv-inv} we find that this potential reads
\begin{align}
V_{\rm MD}(\Phi)
=
\frac{N_s}{64\pi^2}
&\left\{
\left[m^2+\lambda|\Phi|^2+\left(3\xi-\frac12\right)H^2\right]^2
\left[\ln\left(\frac{m^2+\lambda|\Phi|^2+\left(3\xi-\frac12\right)H^2}{\mu^2}\right)-\frac32\right]\right.
\nonumber\\
&+\frac{1}{15}H^4\ln\left(\frac{m^2+\lambda|\Phi|^2+\left(3\xi-\frac12\right)H^2}{\mu^2}\right)
\nonumber\\
&-
\left((1-\delta)\lambda|\Phi|^2+\frac14H^2\right)^2\left[\ln\left(\frac{(1-\delta)\lambda|\Phi|^2+\frac14H^2}{\mu^2}\right)-\frac32\right]
\nonumber\\
&\left.+\frac{59}{240}H^4\ln\left(\frac{(1-\delta)\lambda|\Phi|^2+\frac14H^2}{\mu^2}\right)
\right\},
\label{VCW_MD_SUSY}
\end{align}
for a matter-dominated (MD) period, and
\begin{align}
V_{\rm RD}(\Phi)
=
\frac{N_s}{64\pi^2}
&\left\{
\left[m^2+\lambda|\Phi|^2\right]^2
\left[\ln\left(\frac{m^2+\lambda|\Phi|^2}{\mu^2}\right)-\frac32\right]\right.
+\frac{2}{15}H^4\ln\left(\frac{m^2+\lambda|\Phi|^2}{\mu^2}\right)
\nonumber\\
&-
\left((1-\delta)\lambda|\Phi|^2\right)^2\left[\ln\left(\frac{(1-\delta)\lambda|\Phi|^2}{\mu^2}\right)-\frac32\right]
\left.+{\frac{11}{30}}H^4\ln\left(\frac{(1-\delta)\lambda|\Phi|^2}{\mu^2}\right)
\right\},
\label{VCW_RD_SUSY}
\end{align}
for a radiation-dominated (RD) period. It is not difficult to find formulae, analogous to \eqref{Smin^2_I} and \eqref{DeltaV_SUSY_I}, describing minima of \eqref{VCW_MD_SUSY} and \eqref{VCW_RD_SUSY}. However, as it will be clear later, they are not very important for our analysis.


\section{Evolution of the Peccei-Quinn field during inflation}

There are two processes which determine the evolution of a spectator scalar field $\varphi$ during inflation. One is a classical motion caused by some effective potential, $V(\varphi)$. Second is a random walking caused by quantum fluctuations in (nearly) de Sitter space-time. Such fluctuations take place for fields which are not too heavy, namely the effective mass of $\varphi$ must be smaller than $\frac32H_I$ (see e.g.~\cite{Lyth:2009zz}). Due to the second process the value of the scalar field averaged over regions slightly bigger than the Hubble volume during inflation is a stochastic random variable. Usually we are interested in the probability distribution of such variable, $P({\varphi},t)$. Its evolution is described by the Fokker-Planck equation \cite{Starobinsky:1994bd}
\begin{equation}
\frac{\partial}{\partial t}P(\varphi,t)=
\frac{1}{3H_I}\frac{\partial}{\partial\varphi}\left[V^\prime(\varphi)P(\varphi,t)\right]+\frac{H_I^3}{8\pi^2}\frac{\partial^2}{\partial\varphi^2}P(\varphi,t)\,.
\end{equation}
During inflation any initial probability distribution tends to the equilibrium one equal
\begin{equation}
P_{\rm eq}(\varphi)=C\exp\left(-\frac{8\pi^2}{3}\,\frac{V(\varphi)}{{H_I^4}}\right)\,.
\label{P}
\end{equation}
For simplicity we neglect effects caused by the slow change of the Hubble parameter during inflation \cite{Hardwick:2019uex}. Moreover, we assume that inflation lasted long enough so the final probability distribution may be well approximated by the above equilibrium one.

We are interested in such probability distributions for both, radial and angular, components of the PQ field $\Phi=\frac{1}{\sqrt{2}}Se^{i\theta}$. In the case of $\theta$ this distribution is flat because axion has no non-trivial potential during inflation (it will be generated much later by some non-perturbative effects). The case of the saxion, $S$, is more complicated. Its probability distribution is approximated by \eqref{P} with the effective potential \eqref{VCW_inf_SUSY} provided the effective saxion mass is small enough. So, to check self-consistency of this approach we have to calculate the effective saxion mass in the region close to the minimum of the effective potential where \eqref{P} is maximal. The position of the global minimum of the potential \eqref{VCW_inf_SUSY} in the case of a quasi-supersymmetric spectrum is given by \eqref{Smin^2_I}.
The saxion mass at this minimum is approximately given by
\begin{equation}
m_S^2
\approx
\frac{N_s}{16\pi^2}{\lambda}\left[(3-12\xi){H_I^2}-m^2\right]\ln\left[\frac{(3-12\xi){H_I^2}-m^2}{2\delta \mu^2}\right]\,.
\label{ms2_min}
\end{equation}
Stochastic fluctuations of $S$ during inflation must be taken into account if $m_S^2\lesssim\frac94 H_I^2$ which gives the condition
\begin{equation}
\frac{\lambda N_s}{36\pi^2}
\left[(3-12\xi)-\frac{m^2}{{H_I^2}}\right]\ln\left[\frac{(3-12\xi){H_I^2}-m^2}{2\delta \mu^2}\right]\lesssim 1\,.
\end{equation}
The coupling constant $\lambda$ is very small (in order to avoid too large isocurvature fluctuations which will be discussed later) so typically the above condition is easily fulfilled and the saxion field is light enough to fluctuate during inflation. 
However, those fluctuations are not very large and the saxion field stays relatively close to the minimum of the potential. 
Approximating the potential near the minimum by a quadratic function with the coefficient \eqref{ms2_min} we may estimate the variance of the saxion field as
\begin{equation}
\left.\frac{\Delta S^2}{S^2}\right|_{\rm min,inf}
=
{\cal{O}}\left(\frac{\delta}{\ln\left[\frac{(3-12\xi){H_I^2}-m^2}{2\delta \mu^2}\right]}\right),
\end{equation}
which is small for small $\delta$.

The behavior of the saxion field during inflation in our model with the CW potential \eqref{VCW_inf_SUSY} is quite different from that in models with the usual tree-level potential \eqref{Vlambda4}. The reason is that these potentials differ substantially at large values of the saxion field. The tree-level potential grows as $\lambda_{\Phi}S^4$ so quantum fluctuations are necessary to generate large $S$, at least in some parts of the expanding space. On the contrary, the CW potential with the curvature corrections has a deep minimum at some large value of the saxion field so classical motion is enough to obtain large $S$ practically in the whole space. Some quantum fluctuations may be necessary to move $S$ away from a local minimum at small field values (if such minimum exists and $S$ seats there when inflation begins).

The observed part of the universe some $50\div60$ e-folds before the end of inflation was inside one Hubble volume. Values of the radial and angular components of the PQ field in that Hubble volume were determined by stochastic processes with probabilities given by appropriate distributions of type \eqref{P}. Later, during the last $50\div60$ e-folds of inflation, that region was inflated into many separate Hubble volumes. The average values of $S$ and $\theta$ over those new Hubble volumes -- let us call them $S_i$ and $\theta_i$, respectively -- stay more or less the same until the end of inflation. However, quantum fluctuations produce perturbations so values of the saxion field $S$ and the axion field $\theta$ in different Hubble volumes just after the end of inflation may be slightly different. 
This may lead to isocurvature perturbations at later stages of the universe evolution. Typically, quantum fluctuations of a light scalar field generated during one Hubble time are of order ${H_I}/(2\pi)$. Thus, strong experimental constraints on isocurvature perturbations typically lead to strong upper bounds on the ratio ${H_I}/S_i$. The initial average value $S_i$ is close to $S_{\rm min,inf}$ given by \eqref{Smin^2_I} so this bound may be approximated by\footnote{Such bound depends on details of generation of DM and may be slightly weaker or slightly stronger for a given set of parameters of the model.}
\begin{equation}
\frac{{H_I^2}}{S_{\rm min,inf}^2}
\approx
\frac{\delta\lambda}{(3-12\xi)-\frac{m^2}{{H_I^2}}}\lesssim10^{-8}\,.
\label{isocurv}
\end{equation}
This bound applies to our model if (almost) all DM originates from the PQ field and if its final distribution ``remembers'' fluctuations generated during inflation. As we will see later, in some cases the evolution after inflation may erase information about such fluctuations. 
The condition \eqref{isocurv} is fulfilled if the product $\delta\lambda$ is sufficiently small. Thus, the upper bound on the coupling $\lambda$ is weaker when the spectrum of fields which couple to the PQ field is closer to the supersymmetric limit.

Let us summarize the effect of the evolution of the PQ field during inflation. Just after the end of inflation the average value of the radial (saxion) field, $S_i$, is close to its value at the minimum of the potential during inflation. This value is much bigger than the axion decay constant at late stages of the universe evolution. The average value of the angular (axion) field, $\theta_i$, is chosen stochastically with a flat probability distribution. The last property is typical for models of ``broken PQ symmetry'' in which initial value of the axion field is determined long before inflation and is not substantially changed during inflation.


\subsection{Temperature effects due to Gibbons--Hawking radiation}

Possible temperature contribution to the potential of the PQ field 
during inflation is given by \eqref{thermal_mass} with $T$ equal to the Gibbons-Hawking temperature $T_{GH}={H_I}/(2\pi)$:
\begin{equation}
\frac{\alpha}{24}\left(\frac{{H_I}}{2\pi}\right)^2|\Phi|^2
=\lambda \frac{n_{\rm eff}(T_{GH})}{96\pi^2}{H_I^2}|\Phi|^2\,.
\label{mT_GH}
\end{equation}
This additional mass term may restore the PQ symmetry for some range of parameters for which potential is relatively flat for small values of $|\Phi|$ (as discussed previously). However, it has little effect on the deep minimum present if ${H_I}$ is big enough. In order to restore the PQ symmetry the above positive term should be bigger than at least the depth of the minimum of the potential. Using equations \eqref{Smin^2_I}, \eqref{DeltaV_SUSY_I} and \eqref{mT_GH} we get the following condition on the effective number of particles present in the thermal bath at $T_{GH}$:
\begin{equation}
\frac{n_{\rm eff}(T_{GH})}{N_s}
\gtrsim
\frac32\left[(3-12\xi)-\frac{m^2}{{H_I^2}}\right]\ln\left(\frac{(3-12\xi){H_I^2}-m^2}{2e\delta\mu^2}\right)\,.
\label{neff/Ns}
\end{equation}
At any temperature $n_{\rm eff}$ may be at most equal 
$\frac32N_s$.  
Thus, condition \eqref{neff/Ns} may be fulfilled only if there is a very strong cancellation of the two terms in the square bracket in its r.h.s. Such cancellation is necessary not only to make the r.h.s.~small but also to increase $n_{\rm eff}(T_{GH})$. 
Masses of fermions and scalars to which the PQ field couples during inflation  are given approximately by
\begin{equation}
m^2_{\rm s,f} \approx 
\frac{(3-12\xi){H_I^2}-m^2}{2\delta}\,.
\label{m2sf}
\end{equation}
Without the mentioned above cancellation those particles are practically absent from the thermal bath because for $\delta\ll1$ their masses are much bigger than the Gibbons-Hawking temperature. So, in models with quasi-supersymmetric spectrum the temperature effects due the Gibbons--Hawking radiation are typically negligible.

Another consequence of the fact that during inflation the effective mass $m_{\rm s}$, given by \eqref{m2sf}, is much bigger than ${H_I}$ 
is that scalar fields $\phi_i$ do not fluctuate during inflation. Otherwise non-zero values of $\phi_i$ would have contributed to the effective mass of the PQ field changing very much its evolution.


\section{Evolution of the Peccei-Quinn field after inflation}

Typically just after inflation, because of the Hubble friction, the PQ field $\Phi$ is frozen at its initial value generated during inflation, with the radial component equal $S_i$ and the angular component equal $\theta_i$. The axion field stays frozen for a long time because its mass vanishes at high scales. Evolution of the saxion field begins when the Hubble parameters decreases to about one third of the saxion effective mass. So, it is important to calculate that effective mass.


\subsection{Saxion effective mass and beginning of oscillations after inflation}

Just after inflation the radial component of the PQ field has value $S_i$ close to the position of the minimum \eqref{Smin^2_I} of the potential \eqref{VCW_inf_SUSY} during the final stages of inflation.  This initial value of $S$ is determined by the potential \eqref{VCW_inf_SUSY} but the potential at the early stages of the reheating process after inflation is changed to \eqref{VCW_MD_SUSY}. Not only the form of the potential is different but also the Hubble parameter is different. During inflation it was almost constant and equal ${H_I}$ while after inflation it decreases with the decreasing energy density.

Typically $S_i$ does not correspond to any minimum of the potential after inflation. Nevertheless, the saxion field $S$ not necessarily immediately starts to evolve towards a minimum of the new potential. The evolution begins only when the Hubble friction decreases to a small enough value. To check when this happens we have to calculate the effective mass of $S$ and compare it to the Hubble parameter $H$. So, we should calculate the second derivative of the new potential at the minimum of the old potential. Let us first neglect any possible thermal effects. Using \eqref{Smin^2_I} in the leading order in large $S$ and small $\delta$ we get
\begin{equation}
m^2_{\rm eff}(S_i)
\approx
\frac{\lambda N_s}{32\pi^2}\ln\left(\frac{(3-12\xi){H_I^2}-m^2}{2\delta\mu^2}\right)
\left[\left(3-12\xi\right)\left(3H_I^2- {c} H^2\right)-2m^2\right],
\label{meff}
\end{equation}
{where $c=\frac14$ ($c=0$) during MD (RD) era.} The saxion field stars to oscillate when $H^2=\left(H_i^{(0)}\right)^2\sim \left(m_{\rm eff}/3\right)^2$. Using this condition and \eqref{meff} with $H=H_i^{(0)}$ we obtain
\begin{equation}
\left(H_i^{(0)}\right)^2
\approx
\frac{\lambda N_s}{32\pi^2}\,
\frac{\left[\left(1-4\xi\right)-\frac{2m^2}{9H_I^2}\right]\ln\left(\frac{(3-12\xi){H_I^2}-m^2}{2\delta\mu^2}\right)}
{1+{c}\frac{N_f\lambda}{{24}\pi^2}(1-4\xi)\ln\left(\frac{(3-12\xi){H_I^2}-m^2}{2\delta\mu^2}\right)}\,{H_I^2}\,.
\label{Hi}
\end{equation}
The coupling $\lambda$ is very small so the value of $H_i^{(0)}$ depends very weakly on the parameter $c$. Hence, we will set it to zero in the following. The saxion field starts to oscillate when the value of the Hubble parameter decreases approximately to
\begin{equation}
H_i^{(0)}
\approx
\sqrt{\frac{\lambda N_s}{32\pi^2}
\left[\left(1-4\xi\right)-\frac{2m^2}{9H_I^2}\right]\ln\left(\frac{(3-12\xi){H_I^2}-m^2}{2\delta\mu^2}\right)}\,{H_I}\,.
\label{H_i^0}
\end{equation}
$H_i^{(0)}$ is much smaller than ${H_I}$ because $\lambda$ must be very small. 
Thus, in our model with thermal effects neglected, the saxion field typically starts to oscillate long after the end of inflation 
when the Hubble parameter is some orders of magnitude smaller than it was during inflation.

\begin{figure}
\includegraphics[scale=0.95]{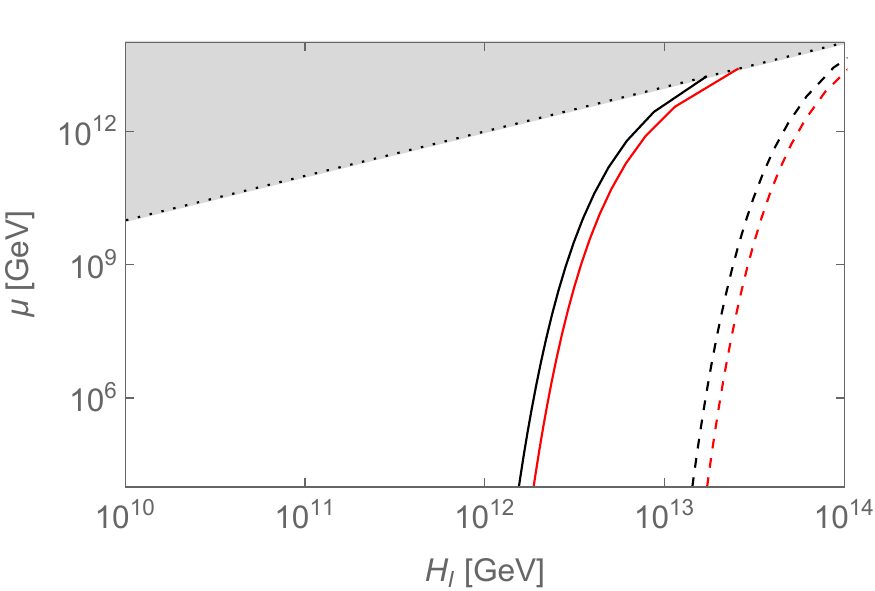}
\hfill
\includegraphics[scale=0.95]{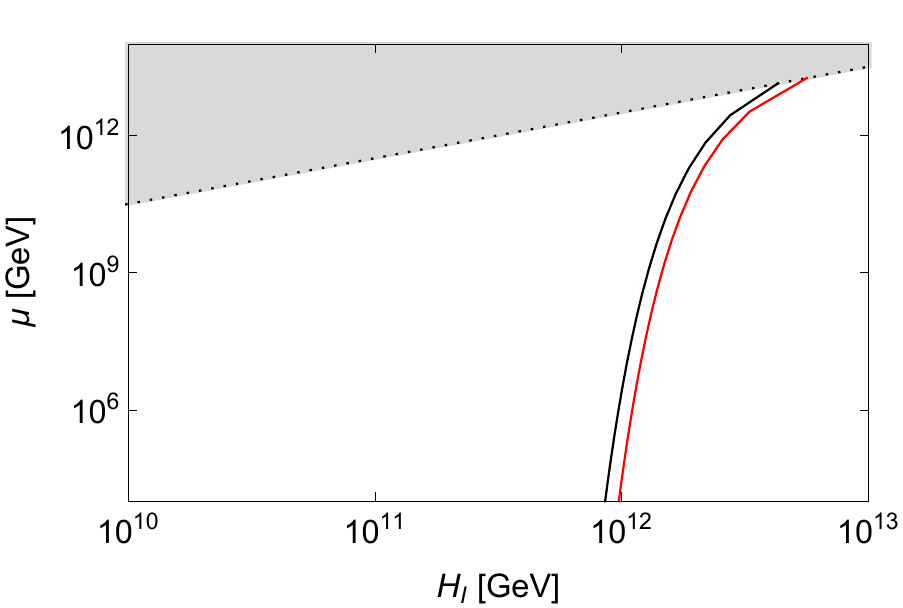}
\caption{Curves for which $H_i^{(0)}=H_i^{(T)}$. 
The black curves are for $m^2$ much smaller than ${H_I^2}$ while the red ones are for $m^2=0.3H_I^2$. 
Left panel: examples with $T_{RH}<\widetilde{T}_{RH}$ and parameters $\lambda=10^{-7}$, $\delta=0.1$, $\epsilon_{RH}=10^{-6}$ (solid curves) and $\epsilon_{RH}=10^{-4}$ (dashed curves), where $\epsilon_{RH}=H_{RH}/H_I$. Right panel: example with parameters $\lambda=10^{-10}$, $\delta=10^{-2}$ 
and $T_{RH}>\widetilde{T}_{RH}$ (this condition is fulfilled for $H_I\gtrsim3(\lambda n_{\rm eff}(\widetilde{T}_{RH})/\epsilon_{RH})10^{15}$\,GeV).
For parameters to the left from a given curve the saxion oscillations start due to the thermal mass while for parameters to the right
they start due to the zero-temperature CW potential.
Our model does not work in the grey region for which there is no minimum at large value of $S$.}
\label{Fig-HT=H0}
\end{figure}

The above estimates of the saxion mass and the value of the Hubble parameter at which the saxion oscillations begin were obtained with thermal effects neglected. Let us now take them into account. Thermal contribution to the saxion mass squared, equal $\frac{1}{24}\alpha T^2=\frac{1}{24}n_{\rm eff}\lambda T^2$, may initiate oscillations if it is bigger than about $(3H)^2$. This happens when the Hubble parameter drops below some critical value $H_i^{(T)}$. The expression for $H_i^{(T)}$ depends on whether the saxion field oscillations start during the reheating process or after it is completed. This, in turn, depends on the reheat temperature. Using the approximation that the universe expands as a matter dominated one during reheating and as a radiation dominated one after reheating we get the following result
\begin{equation}
H_i^{(T)}
\approx
\begin{cases}
\displaystyle\sqrt[6]{\frac{5}{g_*}}\frac{\sqrt[3]{12}}{72\sqrt{\pi}}\left(n_{\rm eff}\lambda\right)^{2/3}T_{RH}^{2/3}M_{Pl}^{1/3}>H_{RH}
& \quad\text{if}\quad {T_{RH}}<\widetilde{T}_{RH}
\\[16pt]
\displaystyle\frac{\sqrt{5}}{144\sqrt{\pi^3g_*}}\,n_{\rm eff}\lambda{M_{Pl}}<H_{RH}
& \quad\text{if}\quad {T_{RH}}>\widetilde{T}_{RH}
\end{cases}
\label{H_i^T}
\end{equation}
where $n_{\rm eff}$ should be calculated at temperature $T_i^{(T)}$ corresponding to $H_i^{(T)}$. 
${T_{RH}}$ and $H_{RH}$ are the values of temperature and Hubble parameter, respectively, at the end of reheating and
\begin{equation}
\widetilde{T}_{RH}=\sqrt{\frac{5\lambda n_{\rm eff}(\widetilde{T}_{RH})}{96\pi^3g_*}}{M_{Pl}}\,.
\end{equation}
In deriving \eqref{H_i^T} we neglected contribution to the effective saxion mass from the zero temperature potential \eqref{VCW_MD_SUSY} or \eqref{VCW_RD_SUSY}. To be more precise one should simultaneously take into account both contributions to the saxion mass, one from the zero temperature potential and another from the thermal corrections. However, usually this is not necessary. When $H_i^{(0)}$ is substantially bigger than $H_i^{(T)}$ then the zero temperature effects are more important and $H_i^{(0)}$ given by \eqref{H_i^0} is a good approximation of the Hubble parameter when the saxion starts to oscillate. In the opposite case, when $H_i^{(T)}$ is substantially bigger than $H_i^{(0)}$, the temperature effects dominate and $H_i^{(T)}$ given by \eqref{H_i^T} is a good approximation of the Hubble parameter when the saxion oscillations start. Or shortly: the saxion oscillations begin when the Hubble parameter is approximately equal
\begin{equation}
H_i \approx \max\left(H_i^{(0)}, H_i^{(T)}\right)\,.
\end{equation}
In order to check which effects initiate saxion oscillations we have to compare $H_i^{(0)}$ and $H_i^{(T)}$. Their ratio is given by
\begin{equation}
\frac{H_i^{(0)}}{H_i^{(T)}}
\approx
3\sqrt[6]{\frac{3g_*}{10}}\sqrt{\frac{3}{\pi}\left[\left(1-4\xi\right)-\frac{2m^2}{9H_I^2}\right]\ln\left(\frac{(3-12\xi){H_I^2}-m^2}{2\delta\mu^2}\right)}\,
\frac{N_s^{1/2}}{n_{\rm eff}^{2/3}\big(T_i^{(T)}\big)}
\left(\frac{{H_I}}{{T_{RH}}}\right)^{\!2/3}
\left(\frac{{H_I}}{\sqrt{\lambda}{M_{Pl}}}\right)^{\!1/3},
\label{H0HT_MD}
\end{equation}
when ${T_{RH}}<\widetilde{T}_{RH}$ and
\begin{equation}
\frac{H_i^{(0)}}{H_i^{(T)}}
\approx
36\sqrt{\frac{\pi g_*}{10}\left[\left(1-4\xi\right)-\frac{2m^2}{9H_I^2}\right]\ln\left(\frac{(3-12\xi){H_I^2}-m^2}{2\delta\mu^2}\right)}
\frac{N_s^{1/2}}{n_{\rm eff}\big(T_i^{(T)}\big)}\,
\frac{{H_I}}{\sqrt{\lambda}{M_{Pl}}}\,,
\label{H0HT_RD}
\end{equation}
when ${T_{RH}}>\widetilde{T}_{RH}$. 
The last two factors in \eqref{H0HT_MD} and the last factor in \eqref{H0HT_RD} show that the relative importance of the zero temperature potential increases with the value of the Hubble constant during inflation. However, there is the upper bound on ${H_I}$ of order $10^{14}$\,GeV. Even for the maximal allowed ${H_I}$ the factor ${H_I}/(\sqrt{\lambda}{M_{Pl}})$ is much smaller than 1 unless $\lambda$ is several orders of magnitude smaller than the limit coming from the bounds on the isocurvature perturbations. Coupling $\lambda$ may be as big as $10^{-8}\delta^{-1}$ so ${H_I}/(\sqrt{\lambda}{M_{Pl}})\sim{\cal{O}}\left(\sqrt{\delta}\,\frac{{H_I}}{10^{15}\,\text{GeV}}\right)$ which typically is many orders of magnitude smaller than 1. In such cases $H_i^{(T)}$ is typically much bigger than $H_i^{(0)}$ so the saxion oscillations start due to the thermal contribution to its mass. This is illustrated by some examples in Fig.~\ref{Fig-HT=H0}. 
$H_i^{(0)}$ may still be bigger than $H_i^{(T)}$ if thermal effects are strongly suppressed by very small $n_{\rm eff}$ at $T_i^{(T)}$. This may happen when particles to which the PQ field couples are practically absent from the thermal plasma (because of too small couplings or too heavy masses). This is not the case for example for the QCD PQ field because such field couples to particles with color charges.
In the examples shown in Figs.~\ref{Fig-HT=H0} and \ref{Fig-AS/S0} we used the effective numbers of degrees of freedom motivated by the QCD axion, namely $N_s=4N_f=12$, $n_{\rm eff}=\frac32N_s=18$.

We showed in this subsection that the effective mass, which initializes the oscillations of the saxion field after inflation, may be dominated by the contribution from the zero-temperature CW potential or by thermal effects. It occurs that the evolution of the PQ field may be quite different in these cases. We will discuss these cases in some detail in the following subsections. But first we will recall results for the case of the simple quartic tree level PQ potential \eqref{Vlambda4} with a small self-coupling presented e.g.~in \cite{Co:2017mop,Co:2020dya,Lyth:1991ub,Lyth:1992tw}.


\subsection{Evolution in tree level potential}

Oscillations of the saxion field in a purely quartic potential lead to production of saxion and axion particles via the parametric resonance mechanism \cite{Kasuya:1996ns,Felder:2000hr,Harigaya:2015hha,Dror:2021nyr}. Numerical simulations are necessary to calculate number densities of produced particles. To estimate the result 
the authors of \cite{Co:2020dya} used the approximation in which 
all the potential energy stored in the saxion field is converted to  similar number of saxions and axions just after the beginning of oscillations. The produced axions are in general relativistic. Axions may become WDM if their momenta are sufficiently redshifted (otherwise they contribute to dark radiation). On the other hand, saxions, which are much heavier, could dominate the total energy density of the universe. Also the saxion oscillations, if they are not totally dumped due to the parametric resonance production of particles, could contribute too much to the energy density. One way to avoid such unacceptable situation is the assumption that the saxions and saxion oscillations are thermalized. In order to achieve this the authors of \cite{Co:2020dya} invoke thermal effect resulting from coupling of the PQ field to the Standard Model via the Higgs portal.


\subsection{Evolution in the CW potential with negligible thermal effects}
\label{sec:evolution_CW}

In models discussed in this work, with different kinds of corrections taken into account, the above presented picture must be modified. In many cases such modifications are very significant. 
First we discuss models in which thermal corrections to the evolution of the PQ field may be neglected so the saxion effective mass which overcomes the Hubble friction comes from the zero temperature CW potential \eqref{VCW_MD_SUSY} or \eqref{VCW_RD_SUSY}. The initial value of the saxion field, $S_i$, is not much different from the position of the minimum \eqref{Smin^2_I} of the CW potential during inflation. When the oscillations begin at time $t_i$ the potential is different. It has different dependence on the Hubble parameter and also the value of this parameter is smaller. The minimum of $\eqref{VCW_MD_SUSY}$ at large field value, if it exists, is located approximately at
\begin{equation}
S_{\rm min,MD}^2
\approx
\frac{(3-12\xi)H^2-4m^2}{4\delta\lambda}\,.
\label{Smin^2_MD}
\end{equation}
The Hubble parameter at $t_i$ is given by \eqref{H_i^0}, so the position of the new minimum at time $t_i$ is at much smaller values of $S$ than the position of the old minimum during inflation. 
When the oscillations start during the reheating process 
this suppression is at least
\begin{equation}
\frac{S_{\rm min,MD}^2}{S_{\rm min,inf}^2}
<
{\cal{O}}\left(\frac{\lambda N_s}{128\pi^2}\right)\,.
\end{equation}
For some ranges of parameters the minimum at \eqref{Smin^2_MD} no longer exists and the only minimum of the potential is close to its late-time position $S_{\rm min,0}$. The above discussion applies also when the saxion field starts to evolve during RD era after the reheating.

It follows from the above discussion that at early stages of oscillations the structure of the potential at small field values (much smaller than the initial amplitude $S_i$) may be treated as small perturbation of the leading term at large field values which is proportional to $\lambda^2S^4\ln(\lambda S^2/\mu^2)$. Oscillations in such potential differ from oscillations in the purely quartic potential $\lambda_\Phi S^4$ applied usually in the literature. The amplitude of oscillations decreases due to the expansion of the universe. In the case of the quartic potential this amplitude scales as the inverse of the cosmic scale factor, ${\cal{R}}$, so the energy density of the oscillating field (sum of potential and kinetic energies averaged over one period of oscillations) decreases as $\rho_{\rm osc}\propto{\cal{R}}^{-4}$. The energy density decreases faster if the potential has an additional logarithmic factor. The dependence of $\rho_{\rm osc}$ on ${\cal{R}}$ in such case changes with time\footnote{One can check that $\rho_{\rm osc}\propto{\cal{R}}^{-k}$ where $k$ is slightly bigger than 4 for very large amplitude and grows with decreasing amplitude, e.g.~$k\to4.5$ when the average of the logarithmic factor in the potential approaches 1.} 
and is always stronger than ${\cal{R}}^{-4}$. Thus, in models with the CW potential the energy of saxion oscillations, so also the energy of particles which can be produced via a parametric resonance, decrease faster than in models with the tree level PQ potential.

When analyzing the resonant production of particles one should take into account also other issues. One of them is related to the structure of the potential at small values of the saxion field. The tree level PQ potential is not purely quartic. It contains also a negative quadratic contribution. It was shown \cite{Kofman:1997yn, Greene:1997fu} that addition of even a small (negative or positive) quadratic term may change very strongly development of the parametric resonance. Production of particles is slower and less effective than in a purely quartic potential. In addition, production of particles stops at all when the amplitude of oscillations decreases to some critical value. Such critical value of the amplitude scales as $\lambda_\Phi^{-1/2}$ so it is quite large in models with the tree level PQ potential with extremely small self-coupling $\lambda_\Phi$.

Another important effect usually neglected in simple analyses is the back-reaction of produced particles on the dynamics of the oscillating field. Analytical estimates \cite{Greene:1997fu} as well as lattice simulations 
\cite{Khlebnikov:1996mc} show that in the case of $\lambda\varphi^4$ theory less than 1\% of energy of oscillations is transferred to the produced particles. Thus, number of particles which may be produced via parametric resonance in such models is much smaller than usually assumed.

Similar arguments apply to models with the CW potential. The number of resonantly produced particles is even smaller due to faster decrease of the amplitude of oscillations. Dedicated numerical simulations are necessary to check how important are the above effects in a given model.

When the amplitude of saxion oscillations is still large, the saxion field vanishes twice during each oscillation period. Thus, the axion angle variable $\theta$ changes its value from $\theta_i$ to $\theta_i+\pi$ and vice versa. 
After some time, when the full temperature-dependent potential develops minima at $S\ne0$ and the amplitude of oscillations sufficiently decreases, the oscillations continue around one minimum of the potential at $S\ne0$ and the value of $\theta$ does not change any more. 
It may be $\theta_i$ or $\theta_i+\pi$, depending on the initial conditions. Anyway, this value, up to the two-fold ambiguity, is determined by stochastic processes during inflation. The same is true for the relic energy density associated with axion oscillations which start later when non-perturbative effects generate axion potential. Isocurvature perturbations of this energy density are also related to inflationary dynamics, and result from fluctuations of $\theta$ around $\theta_i$ generated during last $50\div60$ e-folds of inflation. This is much different from the picture often assumed for axions with non-trivial dynamics of the PQ field after inflation in which misalignment mechanism leads to strong white noise at small scales \cite{Feix:2019lpo,Irsic:2019iff,Feix:2020txt}.


\subsection{Evolution in the CW potential with important thermal effects}
\label{sec:evolution_thermal}

Now we switch to the situation when thermal effects play an important role in the evolution of the PQ field. Such effects may be crucial from the very beginning of this evolution or may become important only later. 
First we discuss the former scenario i.e.~models in which the thermal contribution to the saxion mass is responsible for the onset of oscillations (the latter case we discuss in subsection \ref{sec:heavy}). The saxion field starts to evolve when the Hubble parameter drops to $H_i^{(T)}$ given by eq.~\eqref{H_i^T}. This is the case when the thermal mass evaluated at $S_i$ is substantially bigger than the second derivative of the zero temperature CW potential \eqref{VCW_MD_SUSY} or \eqref{VCW_RD_SUSY}. If the thermal term (which is quadratic in $S$) dominates over the CW potential (which is quartic with additional logarithmic enhancement) at large $S=S_i$ it also dominates at smaller values of $S$. Thus, the saxion starts to oscillate in potential which may be approximated by a quadratic one. This has vary important consequences for the resonance production of saxion and axion particles. 
Namely, such production is very strongly suppressed. Particles may be produced via a parametric resonance when the appropriate adiabaticity condition is violated i.e.~when masses of these particles change fast enough during oscillations \cite{Kofman:1997yn}. But in the (approximately) quadratic potential masses are (approximately) constant\footnote{Some particle production is possible due to corrections to the quadratic potential \cite{Co:2020jtv} but this depends on the form and size of such corrections.}. In principle particles other than saxions and axions, i.e.~scalars $\phi_i$ and fermions $\psi_j$, could be still produced. We will discuss this possibility in subsection \ref{sec:thermalization}.

At early stages of oscillations the potential for the interesting range of $S$ may be approximated by a quadratic one and as a result there is practically no resonant production of saxions and axions. An important question is whether and how this may change later. There are two processes we should take into account. On one hand, the temperature decreases so the value of the thermal mass decreases. On the other hand, the amplitude of oscillations also decreases. This of course lowers available potential energy but the thermal part (quadratic in amplitude) decreases slower than the CW part (depending more strongly than quartically on the amplitude). One has to check how long the thermal mass term keeps its initial dominance over the CW potential.

Process of reheating after inflation begins when the energy density of the universe is strongly dominated by the inflaton oscillations and the universe expands like in a MD era. Then the contribution to the energy from the produced relativistic particles gradually increases and the universe tends to a RD one. One can easily describe saxion oscillations in both limits of MD and RD. The equation of motion for the saxion field when its potential is dominated by the thermal mass is approximated by
\begin{equation}
\ddot{S}+3H\dot{S}+\frac{n_{\rm eff}(T)\lambda}{24}\,T^2S\approx0\,.
\label{eomS}
\end{equation}
We know how the temperature and the Hubble parameter depend on the cosmic scale factor ${\cal{R}}$ during and after reheating:
\begin{equation}
\begin{tabular}{lll}
$T \sim {\cal{R}}^{-3/8}$ &\qquad $H \sim {\cal{R}}^{-3/2}$
&\qquad during reheating (MD)\,,
\\[4pt]
$T \sim {\cal{R}}^{-1}$ &\qquad $H \sim {\cal{R}}^{-2}$
&\qquad after reheating (RD)\,.
\end{tabular}
\end{equation}
Using this in \eqref{eomS} we find that the amplitude of $S$ oscillations, $A_S$, changes as
\begin{equation}
\begin{tabular}{ll}
$A_S \sim {\cal{R}}^{-21/16}$ 
&\qquad during reheating (MD)\,,
\\[4pt]
$A_S \sim {\cal{R}}^{-1}$ 
&\qquad after reheating (RD)\,.
\end{tabular}
\end{equation}
The maximal energy density due to the thermal mass, $\varrho_{\rm th}$, is proportional to $T^2A_S^2$. The maximal energy density of the CW potential, $\varrho_{\rm CW}$, changes logarithmically faster than $A_S^4$. Thus 
\begin{equation}
\begin{tabular}{lll}
$ \varrho_{\rm th}\sim {\cal{R}}^{-27/8}$ &\qquad $\varrho_{\rm CW} \sim {\cal{R}}^{-21/4}\ln{\cal{R}}^{-1}$
&\qquad during reheating (MD)\,,
\\[4pt]
$\varrho_{\rm th} \sim {\cal{R}}^{-4}$ &\qquad $\varrho_{\rm CW} \sim {\cal{R}}^{-4}\ln{\cal{R}}^{-1}$
&\qquad after reheating (RD)\,.
\end{tabular}
\label{rho_MD_RD}
\end{equation}
The maximal energy of the zero temperature CW potential decreases in the expanding universe much (slightly) faster than the maximal energy of the thermally generated mass during (after) reheating. If the thermal mass dominates at the onset of saxion oscillations it dominates even more at later times. As a result, the energy of saxion oscillations scales approximately as $\varrho_{\rm th}$ given in \eqref{rho_MD_RD} and there is no substantial production of axions and saxions via a parametric resonance.

The above conclusions are valid as long as the used approximations apply. In estimating the maximal contribution to the energy density 
coming from the zero temperature CW potential we used its asymptotic behavior for large $S$. This approximation evidently breaks down when the amplitude of oscillations decreases to values for which the full  structure of the potential with its extrema becomes non-negligible. 
Domination of the thermal contribution to the potential ends at temperatures close to the temperature at which the thermal mass is equal to minus the curvature of the CW potential at small values of $S$. Neglecting modifications from the Hubble constant (which may be already relatively small), this temperature is approximately equal
\begin{equation}
\widetilde{T}^2\approx\frac{3}{4\pi^2}\,\frac{N_s}{n_{\rm eff}\big(\widetilde{T}\big)}\,\ln\left(\frac{e\mu^2}{m^2}\right)m^2\,.
\label{Ttilde}
\end{equation}

In order to discuss the evolution of the PQ field at temperatures below $\widetilde{T}$ we need to estimate the amplitude of saxion oscillations at $T=\widetilde{T}$. An important quantity is the ratio of that amplitude to the position of the minimum of the zero temperature CW potential.  Let us consider first the simpler case when the saxion oscillations start after the reheating. Using equations  \eqref{Smin^2_SUSY_0}, \eqref{Smin^2_I}, \eqref{H_i^T}, \eqref{Ttilde} and the fact that during a RD epoch the amplitude changes as $A_S\sim{\cal{R}}^{-1}\sim H^{1/2}$ we get
\begin{equation}
\frac{A_S^2(\widetilde{T})}{S_{\rm min, 0}^2}
\approx
\frac{36\pi g_*}{5}\,
\frac{N_s}{n_{\rm eff}\big(T_i^{(T)}\big)n_{\rm eff}\big(\widetilde{T}\big)}\,
\frac{\frac{m^2}{\mu^2}\ln\left(\frac{e\mu^2}{m^2}\right)}{1-\frac{m^2}{2\mu^2}+\frac{m^4}{24\mu^4}}\,
\frac{(3-12\xi){H_I^2}-m^2}{\delta\lambda M_{Pl}^2}\,.
\label{AS/S0}
\end{equation}
If oscillations start during reheating the r.h.s.~of the above formula should be multiplied by the additional factor of 
\begin{equation}
{\cal{O}}\left({6}\right)\cdot
\left({\frac{{T_{RH}}}{\sqrt{\lambda n_{\rm eff}\big(T_i^{(T)}\big)}{M_{Pl}}}}\right)^{1/3}\,.
\label{AS/S0_extra_factor}
\end{equation}

The value of the ratio ${A_S^2(\widetilde{T})}/{S_{\rm min, 0}^2}$ is very important for further evolution of the PQ field and for the relic abundance of warm and cold axions. One may consider a few cases with the following different qualitative features.

\begin{itemize}
\item[{\bf A:}] {\boldmath${A_S^2(\widetilde{T})}\gg{S_{\rm min, 0}^2}$}

Amplitude of saxion oscillations is still relatively big when the zero temperature CW potential starts to dominate over the thermal mass contribution. The potential is no longer (approximately) quadratic so production of axion and saxion particles via parametric resonance may be possible. However, the energy available for such production is much smaller than in the case with thermal corrections neglected. 
Oscillations triggered by thermal mass begin much earlier so their amplitude is much more redshifted.
Formulae \eqref{rho_MD_RD} show that this effect is especially strong if saxion oscillations are thermally initiated during the reheating process. In addition, the number of produced axions may be very different from the number of produced saxions\footnote{Even in a simple model considered in \cite{Desroche:2005yt} the relative amount of two kinds of particles produced via parametric resonance depends strongly on the ratio of some coupling constants.}.
It is very difficult to predict amounts of resonantly produced particles without dedicated numerical simulations. However, one may expect that the resulting axion contribution to WDM and saxions which must be later thermalized may be substantially smaller than obtained without thermal corrections taken into account.

On the other hand, the axion contribution to CDM should be very similar to the case with negligible thermal mass. The saxion field still oscillates when the ${U(1)_{PQ}}$ symmetry is finally broken i.e.~when the full potential develops 
a minimum close to its final position $S_{\rm min,0}$ given in \eqref{Smin^2_SUSY_0}. The relic abundance of cold axions depends on the initial value of the angular variable, $\theta_i$, determined by stochastic processes during inflation.

\item[{\bf B:}] {\boldmath${A_S^2(\widetilde{T})}\sim{S_{\rm min, 0}^2}$}

Situation is quite similar to the previous case, especially for the axion contribution to CDM. The number of produced saxions and warm axions is rather small. Close to the minimum at $S=0$ the potential may very well be approximated by a quadratic one. As we already mentioned, resonant production of axions and saxions in this approximation is strongly suppressed. The saxion field oscillates in this quadratic potential with energy which redshifts as that of non-relativistic matter. Finally saxions and warm axions are produced due to a tachyonic instability but the available energy is rather small. The resulting warm axion contribution to the energy density of DM is a growing function of ${A_S^2(\widetilde{T})}$. A bigger amplitude of oscillations leads to later particles production when the global minimum of the potential is deeper. In addition, density of later produced particles is less diluted by the expansion of the universe.

\item[{\bf C:}] {\boldmath${A_S^2(\widetilde{T})}\ll{S_{\rm min, 0}^2}$}

Late stages of the PQ field evolution are much different in cases for which the thermal mass dominates long enough that the amplitude of saxion oscillations decreases much below $S_{\rm min, 0}$. When the temperature approaches the critical value $\widetilde{T}$ the full potential develops a minimum at $S\ne0$ with its position moving toward the final value $S_{\rm min, 0}$. 
The later evolution of the PQ field is dominated by a tachyonic instability. 
A case of a complex field with $U(1)$-symmetric potential was investigated in \cite{Felder:2000hj,Felder:2001kt}.
It was shown that such field which initially has very small velocity and is close to the maximum of the potential decays very quickly into corresponding particles. Moreover, if the initial value of this field is small enough the information about its initial phase is largely erased. 
This property is very important for the production of cold axions in our model. 
If the information about the initial angle $\theta_i$ is lost the final angle $\theta$ has different, uncorrelated values in regions which were casually unrelated at the onset of the tachyonic instability. The resulting axion contribution to CDM density has the characteristics of strong (fluctuations are of order 1) white noise at small scales \cite{Feix:2019lpo,Irsic:2019iff,Feix:2020txt}. Bounds from isocurvature perturbations are significantly relaxed. 

Some properties of the potential at temperatures close to $\widetilde{T}$ has very important consequences for the evolution of the PQ field. So, it is useful to consider two subcases.

\item[{\bf C1:}] {\bf no barrier between global minimum and \boldmath$S=0$}

When the temperature drops to values at which the full potential develops the minimum at $S\ne0$ the saxion field is close to the local maximum of that potential at $S=0$. 
In this case the amounts of warm axions and saxion particles produced due to a tachyonic instability as well as the energy of residual saxion oscillations are all very small. The reason is that, contrary to models investigated in \cite{Felder:2000hj} and \cite{Felder:2001kt}, the potential in our model changes with time due to its dependence on temperature (and Hubble constant if it is still non-negligible at temperature $\widetilde{T}$). Thus, when the minimum of the potential at $S\ne0$ develops it is very shallow. 
The PQ field moves quite quickly to this minimum because of the tachyonic instability but the energy available for production of particles is very small. 
Later the PQ field follows the changing position of the minimum of the potential.

\item[{\bf C2:}] {\bf barrier between global minimum and \boldmath$S=0$ (for some temperatures)}

For some values of the parameters and some range of temperatures the full potential has a barrier separating the global minimum from the region of small values of $S$. In case C the amplitude of saxion oscillations at such temperatures is small so because of this barrier the saxion field keeps oscillating around $S=0$ with a small amplitude despite the fact that there is a deeper minimum at bigger values of $S$. However, the height of such barrier decreases with decreasing temperature and at some point becomes smaller than the energy of saxion oscillations so the saxion field crosses the barrier and evolves towards the global minimum of the potential. Similarly as in case C1, a tachyonic instability is responsible for quick production of saxions and warm axions. The important difference with respect to case C1 is that now the global minimum of the full potential has a non-negligible depth so non-negligible number of particles may be produced. 
Particles produced this way may contribute more to the present density of DM if the potential barrier is crossed later. 
There are two reasons. First, the depth of the global minimum, so also the energy available for the particle production, increases with time. Second, later production of particles results in less dilution caused by the expansion of the universe. Thus, in this case the contribution of warm axions to the present density of DM is a decreasing function of the initial amplitude of saxion oscillations. This dependence on the initial conditions is opposite to that in the case B.

\end{itemize}

Scenario C is somewhat similar to that investigated in \cite{Harigaya:2019qnl}. However, there are two important differences. In the model considered in \cite{Harigaya:2019qnl} the saxion field has vanishing value before the phase transition which leads to white noise fluctuations of the cold axion density after the phase transition. In our model the saxion field oscillates with some fixed value of the PQ phase which (unless the amplitude of such oscillations becomes extremely small) may result in much smaller fluctuations of the density of the cold axions. Second difference is even more important. The thermal effect considered in \cite{Harigaya:2019qnl} modify the PQ potential only at very small values of $S$. In our model thermal corrections are important also for bigger values of $S$ which gives a more complicated time-dependent full potential. Evolution of the PQ field in such time-dependent potential leads to several interesting features discussed in this paper.

Distribution of axion CDM in case C may be very different from cases A and B if the amplitude of the saxion oscillations decreases to very small values before the PQ phase transition and the information of the initial phase of the PQ field is largely erased due to quantum fluctuations. 
We would like to know when such scenario leading to white noise fluctuations may be realized in our model. 
So, we need to identify regions of the parameter space for which the ratio \eqref{AS/S0}, possibly multiplied by \eqref{AS/S0_extra_factor}, may be very small.
Let us start with models when the saxion field oscillations start after reheating and $n_{\rm eff}$ is not very small. In such situation the product of two first factors in \eqref{AS/S0} may be of order $10^{1\div2}$. The third factor is of order 1 for $m^2/\mu^2={\cal{O}}(1)$ and scales as $m^2/\mu^2$ for small $m^2$.\footnote{The third factor in \eqref{AS/S0} is of order 10 for $m^2$ close to the maximal allowed value $e\mu^2$. This follows if one replaces the approximate expression in the denominator of that factor by the exact one calculated numerically. However, this possibility is not very interesting because it requires fine tuning of parameters and in addition in such limit the position of the minimum of the late time potential decreases to zero.
} 
If there are no strong cancellations in the numerator of the last factor in \eqref{AS/S0}, that factor is of order ${H_I^2}/(\delta\lambda{M_{Pl}^2})$. Neglecting in addition the logarithmic correction we may roughly estimate the order of magnitude of the ratio in question:
\begin{equation}
\frac{A_S(\widetilde{T})}{S_{\rm min, 0}}
\approx
{\cal{O}}\left(\frac{1}{\sqrt{\delta\lambda}}\,\frac{m}{\mu}\,\frac{{H_I}}{10^{18}\,\text{GeV}}\right)\,.
\label{AS/S0_estimate}
\end{equation}
There is an upper bound on the product $\delta\lambda$ of order $10^{-8}$. So, scenario C may lead to white noise fluctuations of the axion CDM when the ratio \eqref{AS/S0_estimate} is many orders of magnitude below 1 (dedicated lattice calculations would be necessary to obtain more quantitative conditions).
So, it may be realized only if $\frac{m}{\mu}{H_I}\ll10^{14}\,$GeV. These conditions become even stronger when $\delta\lambda$ is smaller than the maximal allowed value. If saxion field starts to oscillate during the reheating process the ratio ${A_S(\widetilde{T})}/{S_{\rm min, 0}}$ may be somewhat smaller due to the factor \eqref{AS/S0_extra_factor} if the reheat temperature is much smaller than $\sqrt{\lambda n_{\rm eff}\big(T_i^{(T)}\big)}{M_{Pl}}$.

\begin{figure}
\includegraphics[scale=0.87]{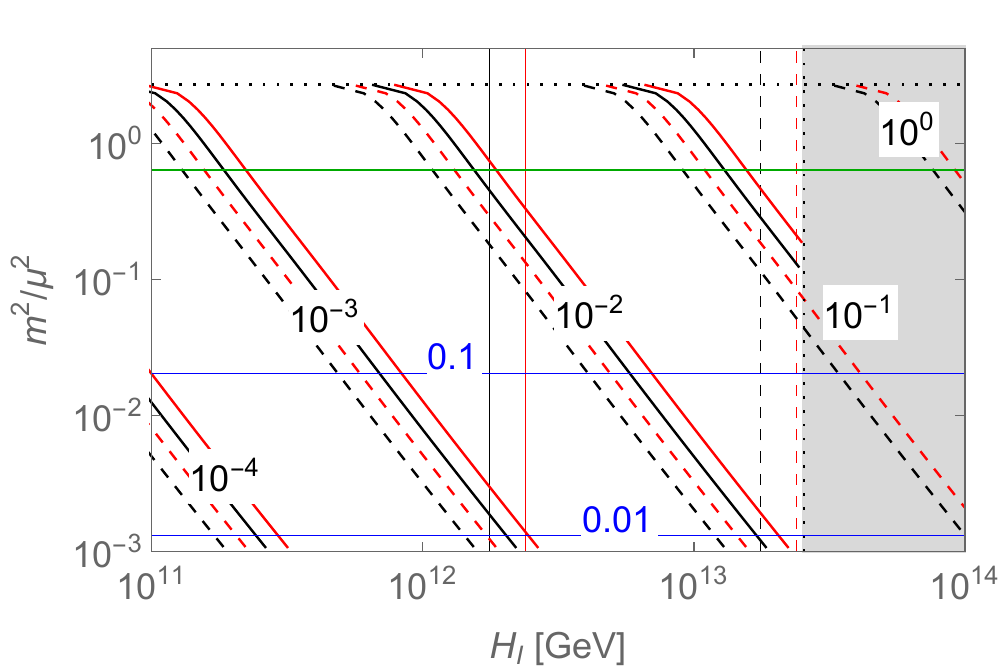}
\hfill
\includegraphics[scale=0.86]{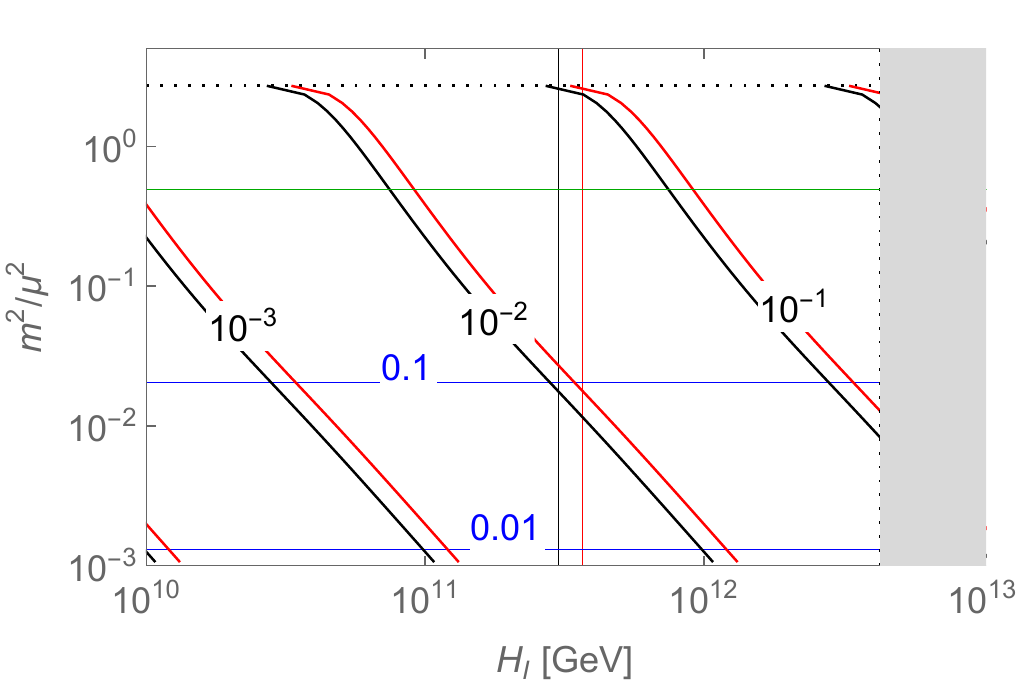}
\caption{Curves (black and red) of constant ratio ${A_S(\widetilde{T})}/{S_{\rm min, 0}}$. Corresponding thin vertical lines denote values of $H_I$ for which masses \eqref{m2sf} are equal to $T_i^{(T)}$ (see discussion in section \ref{sec:heavy}). Parameters used for black and red lines are panel-wise the same as in Fig.~\ref{Fig-HT=H0}.
Constant values of the expression $(m^2/\mu^2)\ln(e\mu^2/m^2)$ are marked by blue horizontal lines. 
Dotted horizontal lines denote the maximal possible value of $m^2/\mu^2$ equal $e$. 
In the grey region there is no minimum at large value of $S$ for any value of $\mu$.}
\label{Fig-AS/S0}
\end{figure}

Some examples illustrating how the ratio ${A_S(\widetilde{T})}/{S_{\rm min, 0}}$ depends on the parameters of the model are shown in Fig.~\ref{Fig-AS/S0}. The left (right) panel corresponds to models with the saxion oscillations starting during a MD (RD) era. As one can see scenario B may be realized typically only when the parameters $H_I$ and $m^2/\mu^2$ are close to their maximal allowed values. Scenario C requires smaller values of at least one of these parameters (there is a smooth transition between both scenarios). If the ratio $m^2/\mu^2$ is too small scenario D of early thermalization takes place (see the subsection \ref{sec:thermalization}). The left panel shows how the results depend on the effectiveness of the reheating process which we parameterize by $\epsilon_{RH}=H_{RH}/H_I$. For smaller reheat temperature $H_I$ or $m^2/\mu^2$ must be bigger to get the same value of ${A_S(\widetilde{T})}/{S_{\rm min, 0}}$ but the dependence on $\epsilon_{RH}$ is not very strong. Results depend also to some extend on the ratio $m^2/{H_I^2}$ (and on the parameter $\xi$ -- this is not shown in the figure but may be easily estimated from eq.~\eqref{AS/S0}). It is possible to obtain bigger values of the ratio ${A_S(\widetilde{T})}/{S_{\rm min, 0}}$ for a given ${H_I}$ by taking much smaller $\delta\lambda$. However, as is shown in the right panel of Fig.~\ref{Fig-AS/S0}, at the same time the maximal allowed value of ${H_I}$ decreases (compare with the right panel of Fig.~\ref{Fig-HT=H0}).
So, it seems that scenario A is the least natural while scenario C is the most natural.

\subsubsection{Saxion oscillations initiated by non-thermal effects}
\label{sec:heavy}

So far in this section we discussed models in which saxion oscillations are initiated by thermal contribution to the saxion mass. Now we move to models in which these oscillations start due to the effective saxion mass dominated by the contribution from the CW potential. This happens when $H_i^{(T)}$ given by \eqref{H_i^T} is smaller than $H_i^{(0)}$ from eq.~\eqref{H_i^0}. One may distinguish two different scenarios leading to such situation\footnote{The third scenario in which thermal effects may be totally neglected was discussed in section \ref{sec:evolution_CW}.}. First: in some regions of the parameter space $H_i^{(T)}<H_i^{(0)}$ even for maximal possible value of $n_{\rm eff}$ of order $N_s$. 
Some examples are shown in Fig.~\ref{Fig-HT=H0} where $H_i^{(T)}<H_i^{(0)}$ in regions to the right from a given curve and it was assumed that $n_{\rm eff}=\frac32N_s$. 
The second scenario is more complicated. $H_i^{(T)}$ is proportional to some positive power of $n_{\rm eff}$. One should remember that this effective number of degrees of freedom may change with time. It depends not only on the couplings of the $\phi$ and $\psi$ fields to the thermal bath but also on their masses. 
The contribution from a given particle to $n_{\rm eff}$ is a monotonically decreasing function of the ratio of its mass to temperature. Such contribution drops substantially below its maximal value when the mass is a few times bigger then $T$ and is negligible when the mass is much bigger that $T$.
Soon after the end of inflation temperature takes its maximal value and then decreases monotonically because of the expansion of the universe. On the other hand, masses of $\phi$ and $\psi$ change in a more complicated way. Equations \eqref{MphiMpsi} show that they depend on a temporary value of the saxion field. Thus, during saxion oscillations $m_\phi$ and $m_\psi$ change non-monotonically. We assume, as e.g.~in \cite{Mukaida:2012qn}, that the time scale of thermal processes is much shorter than the time scale of (saxion) oscillations. So, contribution of $\phi$ and $\psi$ particles to $n_{\rm eff}$ depends on the value of $S$ at given time.

We consider now the situation when masses of $\phi$ and $\psi$ just after inflation (i.e.~evaluated for the saxion field equal to its initial  value $S=S_{\rm min,inf}$ given by \eqref{Smin^2_I}) are too large to produce thermal effects big enough to initialize saxion oscillations. However, when the oscillations are triggered by some other mechanism, thermal effects become important whenever the value of $S$ is small enough. Resonant production is possible when an appropriate adiabaticity condition is violated. Typically it happens when the oscillating field goes through the minimum of its potential. But this is exactly this part of the oscillation cycle when the mentioned above thermal effects are relatively more important. Thus, the effective saxion potential close to the minimum still may be dominated by the thermal contribution which is quadratic in the saxion field. Recalling that there is no parametric resonance in purely quadratic potential we conclude that resonant production of axions and saxions may be significantly affected even when thermal effects do not dominate at large values of the oscillating field. This important conclusion is valid not only for our model with the zero-temperature potential given by \eqref{VCW_MD_SUSY} or \eqref{VCW_RD_SUSY} but also for models with the usually assumed simple tree-level potential \eqref{Vlambda4}. However, numerical calculations are necessary to check how much densities of produced particles are changed due to this effect in a given model.

The strength of the presented here mechanism suppressing the resonant production of particles may be estimated by using the known results for the quartic potential. It was shown in \cite{Greene:1997fu} that a resonant production of quanta of a field oscillating in a quartic potential with a positive quadratic addition is not possible if the amplitude of oscillations drops below the square root of the ratio of coefficients of the quadratic and quartic terms. Applying this result to our model we expect that the resonant production of axions and saxions stops when the amplitude of saxion oscillations drops to value of order 
\begin{equation}
A_S\sim\sqrt{\frac{4\pi^2 n_{\rm eff}}{3 \delta\lambda N_s}}\, T\,,
\end{equation}
where $n_{\rm eff}$ should be calculated for temperature $T$ and small value of $S$. 
In some cases the above critical amplitude may be even as large as the initial amplitude \eqref{Smin^2_I} so the resonant production does not take place at all. To have any resonant production $S_{\rm min,inf}$ must be bigger than the r.h.s.~of the above formula with $T$ replaced with $T_i^{(0)}$ corresponding to $H_i^{(0)}$ given by \eqref{H_i^0}. For example when the saxion field starts to oscillate during RD era and $m^2\ll H_I^2$ this condition may be approximated by
\begin{equation}
H_I \gtrsim \sqrt{\lambda}{M_{Pl}}\sqrt{\frac{5N_s\ln\left[(3-12\xi)H_I^2/(2\delta\mu^2)\right]}{32\pi g_*(1-4\xi)}}\,.
\end{equation}
For the parameters used in the right panel of Fig.~\ref{Fig-AS/S0} the r.h.s.~of the above formula is at least of order $1.6\cdot10^{13}$\,GeV. Thus, there is no substantial resonant production of axions and saxions even if the oscillations start due to the effective saxion mass coming from the CW potential. Slightly more complicated but analogous calculations indicate that the last conclusion applies also to examples shown in the left panel of Fig.~\ref{Fig-AS/S0}.

Regions of the parameter space to which the approximate analytical formulae presented in this section best apply are indicated in Fig.~\ref{Fig-AS/S0} by vertical lines. Each such line shows the value of $H_I$ for which $\phi$ and $\psi$ masses evaluated at $S=S_{\rm min,inf}$ are equal to the corresponding $T_i^{(T)}$ obtained with the assumption that $n_{\rm eff}=\frac32N_s$. These masses increase with increasing $H_I$ so the farther to the right of a given vertical line we go the 
bigger is the ratio of these masses to the temperature at which oscillations start. Thus, the thermal effects become weaker than those obtained with the approximate equations \eqref{thermal_mass}--\eqref{alpha}. This means that the curves of constant ${A_S(\widetilde{T})}/{S_{\rm min, 0}}$ shown in that figure should be somewhat modified for values of $H_I$ substantially bigger than those indicated by appropriate vertical lines. There are two competing effects leading to such modifications. When $\phi$ and $\psi$ masses are big
the oscillations start later (when $H\approx H_i^{(0)}$) which tends to increase the final amplitude. On the other hand, the amplitude of oscillations decreases faster when the potential depends on a higher power of the field. This is one of situations when any quantitative predictions may be obtained only from detailed numerical calculations using the full (not only approximate) thermal correction to the potential. Results of some of such calculations are presented in the next subsection.


\subsection{Relic density of warm and cold axions}
\label{sec:numerical_results}

So far we investigated how different kinds of corrections (radiative, geometric and thermal) may change the evolution of the PQ field during and after inflation. We discussed also some qualitative features of the production of warm and cold axions. 
In this subsection we study in more detail how such corrections 
change the relic density of axions. We illustrate the general features of the model with several quantitative results obtained with numerical calculations.

\begin{figure}
\begin{center}
\includegraphics[scale=0.95]{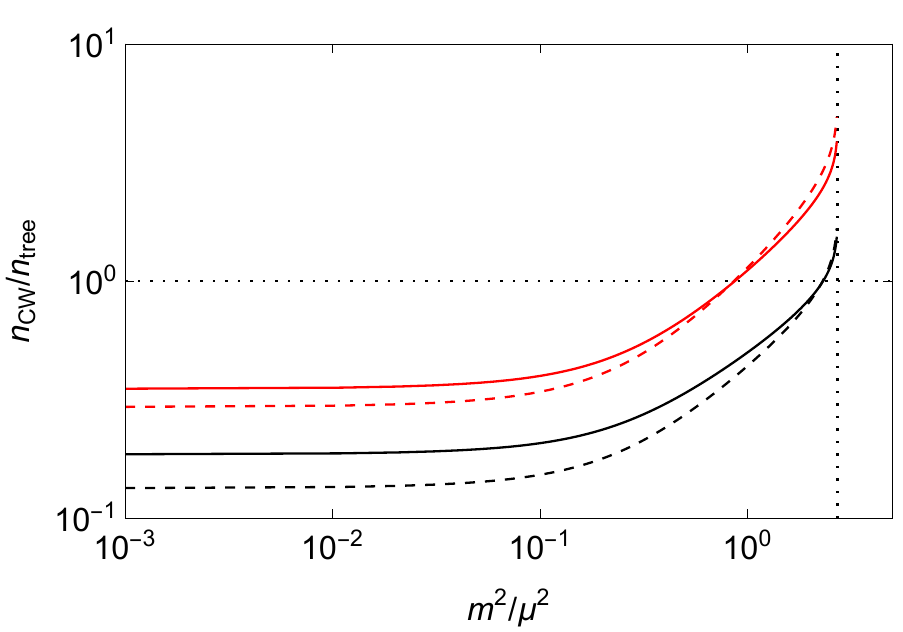}
\hfill
\includegraphics[scale=0.95]{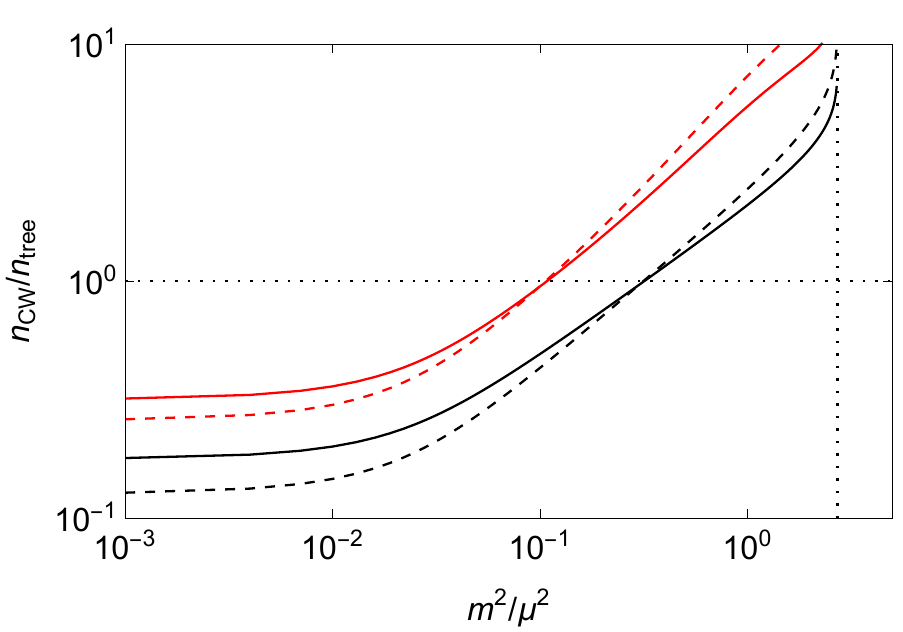}
\end{center}
\caption{The ratio of densities of warm axions produced via a parametric resonance with ($n_{CW}$) and without ($n_{tree}$) radiative corrections taken into account as a function of $m^2/\mu^2$. The relevant parameters are fixed as: $\delta=0.1$ (left panel) and $\delta=0.01$ (right panel); 
$H_I/\mu=5$ (red curves) and $H_I/\mu=10^3$ (black curves). The solid (dashed) lines correspond to situation when the axions are produced after (before) the end of the reheating process.
}
\label{Fig-V0-VCW}
\end{figure}

First we investigate the impact of the radiative corrections on the relic density of warm axions. We want to compare number densities of axions produced via a parametric resonance during saxion oscillations in two cases. One with the simple tree level potential \eqref{Vlambda4} and another in which radiative corrections lead to the Coleman-Weinberg potential \eqref{VCW}. It is not obvious how to compare such two models because they have quite different sets of parameters. We use the following procedure. For given values of the parameters of the model with the CW potential we find such parameters ($\lambda_\Phi$ and $f_a$) of the tree level potential \eqref{Vlambda4} that the axion decay constant and the saxion mass are the same in both models. Next, using the Fokker-Planck probability distribution \eqref{P}, we calculate for both models the most probable initial values of the saxion field generated during inflation. Then we estimate the number density of the produced axions. We apply the approximation (used e.g~in \cite{Co:2019jts}) in which a parametric resonance leads to the production of similar number of axions and saxions, each equal one half of the ratio of the energy of the saxion field to its mass calculated shortly after the onset of saxion oscillations. Finally we calculate the ratio of the axion densities 
obtained in both models rescaled to a common time (in a model in which the saxion field starts to oscillate earlier the number density of produced particles is more diluted due to the expansion of the universe).
This ratio, $n_{CW}/n_{tree}$, depends on three dimensionless parameters:  $m/\mu$, $\delta$ and $H_I/\mu$. Its value is bigger for bigger $m/\mu$ and for smaller $\delta$ and $H_I/\mu$. It occurs that taking the radiative corrections into account may result in smaller or bigger number of warm axions. However, the change usually is not bigger than a factor of a few (it is larger only for $\delta\lesssim{\cal{O}}(0.01)$ and not very small $m/\mu$). Such changes are slightly bigger if the saxion oscillations start before the end of reheating. Radiative corrections modify also the warmness of the produced axions but the modifications are even smaller than in the case of densities (and have opposite dependence on the parameters).
The numerical results for some sets of parameters are shown in Fig.~\ref{Fig-V0-VCW}.

Now we discuss how the results obtained for the model with the CW potential \eqref{VCW} are modified by the geometric and thermal corrections. We use the following procedure. We numerically calculate number densities of warm axions for four cases: without corrections, with only geometric corrections, with only thermal corrections and with both types of corrections taken into account. The results are denoted by: $n_{CW}$, $n_{CW+G}$, $n_{CW+T}$ and $n_{CW+T+G}$, respectively.
Two different approximations are used to estimate amounts of produced warm axions. In the case of a parametric resonance production (when thermal corrections are neglected or are small) we use the same approximation as in the analysis described in the previous paragraph. If the thermal corrections are strong enough there is no resonant production of particles (the full potential is approximately quadratic for saxion field values relevant for the resonance). In such cases saxions and axions may be produced via a tachyonic instability and the number density of produced particles is approximated by the ratio of the available potential energy to the saxion mass evaluated at the minimum of the potential (this approximation was used e.g.~in \cite{Harigaya:2019qnl}).

The main effect of the geometric corrections, as described in section \ref{sec:G_inf}, is a change of the value of the saxion field at the onset of oscillations. Typically this initial value of the saxion field is much bigger than in the case without corrections. Such change of the initial amplitude of the saxion oscillations results in a change of the number of produced warm axions.

The analysis of the thermal corrections is more involved. These corrections modify the potential for the PQ field in a time-dependent way because the temperature decreases with time. The evolution of the PQ field in such time-dependent potential is rather complicated so numerical calculations are necessary to obtain any quantitative results. 
So far, in the  qualitative discussions of thermal corrections, we used the approximation \eqref{thermal_mass}--\eqref{alpha} with the effective number of degrees of freedom \eqref{neff}. 
In our numerical calculations we use the full thermal correction to the potential
\begin{equation}
V_T(\Phi)=\frac{T^4}{2\pi^2} \left[\sum_{\rm bosons}  J_+\left(\frac{M_{\phi_i}}{T}\right)
+4\sum_{\rm fermions}  J_-\left(\frac{M_{\psi_j}}{T}\right)\right]
\end{equation}
where
\begin{equation}
J_{\pm}(y)=\pm\int_0^\infty x^2 \ln\left[1\mp\exp\left(-\sqrt{x^2+y^2}\right)\right]{\rm d} x
\end{equation}
with masses $M_{\phi_i}$ and $M_{\psi_j}$ given by eqs.~\eqref{MphiMpsi}.

We have performed numerical calculations for many sets of values of the parameters of our model. For each set of parameters we numerically integrated the equations of motion for the saxion field and for its quantum fluctuations and calculated the time-dependent number density of saxions. Then we were inspecting the time dependence of the obtained number densities looking for some features characteristic for the parametric resonance \cite{Kofman:1997yn}, \cite{Greene:1997fu}. 
As a cross-check we calculated also whether and how strongly the adiabaticity condition was violated. Our results show that indeed there is no resonant production of particles when the thermal corrections are strong enough i.e.~when the condition $H_i^{(T)}\gtrsim H_i^{(0)}$ is satisfied.

For a big part of the parameter space of our model the thermal corrections are strong enough to suppress any resonant particle production. In such situation warm axions may be produced due to a tachyonic instability. Further numerical calculations are necessary to estimate the resulting number density of such axions. 
First one has to find the amplitude of the saxion oscillations at the time of the phase transition in order to know which of the scenarios -- A, B or C discussed in section \ref{sec:evolution_thermal} -- is realized. In most cases it was scenario C. In order to distinguish between sub-scenarios C1 and C2 one has to check whether the full potential develops at any temperature a barrier between $S=0$ and the minimum. Only with such a barrier the scenario C2 may be realized. Then it is necessary to follow the late evolution of the saxion field in order to find the moment of tachyonic instability and the shape of the potential at that moment. The latter determines the number density of the produced particles while the former is important for the later dilution 
of this density caused by the expansion of the universe.

The resulting number density of warm axions depends on all the parameters of our model. Moreover, such dependence is different for different approximations i.e.~when geometric or thermal corrections are neglected or taken into account. Thus, it would be quite difficult to perform and present a very detailed scan of the whole parameter space of the model. So, instead of doing such scan we will discuss the main features and illustrate them using numerical results for several characteristic benchmark points in the parameter space. The results for these benchmark points are presented in table \ref{table-BP}. For simplicity we will concentrate on the situation when the saxion oscillations start after the reheating is completed.

Let us start our analysis with cases for which the thermal corrections are neglected. This is relatively easy because the corresponding axion densities $n_{CW}$ and $n_{CW+G}$ depend mainly on the initial value of the saxion field $S_i$ when the saxion oscillations begin. Using the know behavior of the CW potential at large field values one can calculate that the initial axion density in the leading approximation equals
\begin{equation}
n_i\approx\frac{1}{32\pi}\sqrt{\frac{N_s\delta}{6}\ln\left(\frac{\lambda S_i^2}{2e\mu^2}\right)}\,\lambda S_i^3\,.
\end{equation}
In order to compare cases with different time of axion production one has to take into account differences of dilution caused by the expansion of the universe. It is convenient to express the number densities in units of the temperature at which we compare different cases. The result reads
\begin{equation}
n_{CW},\,n_{CW+G}
\approx
\frac{1}{12}
\left[\frac{4\pi^3g_*}{5}\right]^{3/4}\left[\frac{3N_s\delta}{2\pi^2}\ln\left(\frac{\lambda S_i^2}{2e\mu^2}\right)\right]^{-1/4}\,\lambda^{-1/2}\left(\frac{S_i}{M_{Pl}}\right)^{3/2}T^3\,.
\end{equation}

In the case with (without) geometric corrections, the initial amplitude of the saxion field is proportional to $\delta^{-1/2}\lambda^{-1/2}H_I$ ($\delta^{-1/4}\lambda^{-1/2}H_I$), so
\begin{equation}
n_{CW} \propto \delta^{-5/8}\lambda^{-5/4}H_I^{3/2}T^3\,,
\qquad\qquad
n_{CW+G} \propto \delta^{-1}\lambda^{-5/4}H_I^{3/2}T^3\,,
\label{n-without-T}
\end{equation}
with the proportionality coefficient bigger in the case of $n_{CW+G}$. Thus, the density is bigger when the geometric corrections are taken into account. The ratio $n_{CW+G}/n_{CW}$ is typically ${\cal{O}}(5)$ for $\delta={\cal{O}}(0.1)$ and grows with decreasing value of $\delta$, e.g.~it is ${\cal{O}}(30)$ for $\delta={\cal{O}}(0.001)$ -- see benchmark points P${}_5$, P${}_8\div$P${}_{11}$ in table \ref{table-BP}. The dependence on the parameters $m$ and $\mu$ is very weak\footnote{In the cases with the geometric corrections $n_{CW+G}$ shows additional dependence on $\xi$ and stronger dependence on $m$ when $m$ is comparable to $H_I$ -- see eq.~\eqref{Smin^2_I}.} -- see benchmark points P${}_4\div$P${}_7$.

The thermal corrections occur to be more important than the geometric ones and may change the relic density of warm axions by several orders of magnitude. However, it is difficult to describe such effects by approximate analytical expressions so numerical calculations are crucial in obtaining quantitative results. First of all, for a given set of parameters one should check which type of the evolution of the saxion field introduced in section \ref{sec:evolution_thermal} (A, B or C) is realized. Scenarios B and especially A require a very small value of the product of parameters $\delta\lambda$ and large values of $H_I$ (see figure \ref{Fig-AS/S0} and discussion below eq.~\eqref{AS/S0_estimate}). So, much more natural is scenario C. It depends on the details of the full time-dependent potential whether it is sub-scenario C1 or C2. Numerical calculations show that the most important from this point of view is the ratio of parameters $m/\mu$. Case C1 is realized for $m/\mu$ bigger than some number of order 0.5 with its precise value depending slightly on $\delta$ (bigger $m/\mu$ required for bigger $\delta$). 
For example, benchmark points P${}_6$, P${}_{10}$ show that the transition region between C1 and C2 (relatively small but non-zero values of $n_{CW+T}$ and $n_{CW+T+G}$) corresponds to $m/\mu\approx0.7$ for $\delta\approx0.1$ and to $m/\mu\approx0.5$ for $\delta\approx0.01$ (while  $m/\mu\approx0.5$ is well in the region C2 for $\delta={\cal{O}}(0.1)$ -- see point P${}_5$). 
The borders between regions leading to scenarios C1 and C2 are shown in Fig.~\ref{Fig-AS/S0} as horizontal green lines. In scenario C1 (above those green lines) a very small amount of warm axions is produced (in the used approximations it is just zero -- see benchmark point P${}_7$). Below the green lines we move smoothly to type C2 scenario with the number of warm axions produced via a tachyonic instability growing quite quickly with decreasing value of the ratio $m/\mu$ (see benchmark point P${}_4\div$P${}_6$).

\begin{table}[t]
\begin{center}
\begin{tabular}{|c||c|c|c|c|c||c|c|c|c|}
\hline
&$\lambda$ & $\delta$ & $m/\mu$ & $\mu$\,[GeV] & $H_I$\,[GeV] &$n_{CW}$ & $n_{CW+G}$ & $n_{CW+T}$ & $n_{CW+T+G}$ \\
\hline\hline
P${}_1$ & $10^{-7}$ & $0.1$ & $0.1$ & {\boldmath$10^{9}$} & {\boldmath$10^{11}$} & $0.042$ & $0.20$ & $4.57\cdot10^5$ & $4.57\cdot10^5$\\
\hline
P${}_2$ & $10^{-7}$ & $0.1$ & $0.1$ & {\boldmath$10^{10}$} & {\boldmath$10^{13}$} & $34$ & $195$ & $4.33\cdot10^5$ & $3.18\cdot10^5$\\
\hline
P${}_3$ & $10^{-7}$ & $0.1$ & $0.1$ & {\boldmath$10^{12}$} & {\boldmath$10^{13}$} & $56$ & $223$ & $4.28\cdot10^5$ & $3.04\cdot10^5$\\
\hline\hline
P${}_4$ & $10^{-7}$ & $0.1$ & {\boldmath$0.1$} & $10^{11}$ & $10^{13}$ & $41.6$ & $206$ & $4.3\cdot10^5$ & $3.1\cdot10^5$\\
\hline
P${}_5$ & $10^{-7}$ & $0.1$ & {\boldmath$0.5$} & $10^{11}$ & $10^{13}$ & $41.9$ & $207$ & $5.0\cdot10^3$ & $2.2\cdot10^3$\\
\hline
P${}_6$ & $10^{-7}$ & $0.1$ & {\boldmath$0.7$} & $10^{11}$ & $10^{13}$ & $42.0$ & $207$ & $95$ & $5.9$\\
\hline
P${}_7$ & $10^{-7}$ & $0.1$ & {\boldmath$0.8$} & $10^{11}$ & $10^{13}$ & $42.0$ & $207$ & $0$ & $0$\\
\hline\hline
P${}_8$ & $10^{-7}$ & {\boldmath$0.03$} & $0.5$ & $10^{11}$ & $10^{13}$ & $84$ & $6.9\cdot10^2$ & $1.9\cdot10^3$ & $3.0\cdot10^3$\\
\hline
P${}_{9}$ & $10^{-7}$ & {\boldmath$0.01$} & $0.5$ & $10^{11}$ & $10^{13}$ & $1.6\cdot10^2$ & $2.0\cdot10^3$ & $1.2\cdot10^3$ &$8.9\cdot10^3$\\
\hline\hline
P${}_{10}$ & $10^{-6}$ & {\boldmath$0.01$} & $0.5$ & $10^{11}$ & $10^{13}$ & $9.0$ & $120$ & $490$ & $47$\\
\hline
P${}_{11}$ & $10^{-6}$ & {\boldmath$0.001$} & $0.5$ & $10^{11}$ & $10^{13}$ & $36$ & $1.1\cdot10^3$ & $370$ & $940$\\
\hline\hline
P${}_{12}$ & {\boldmath$10^{-7}$} & $0.1$ & $0.2$ & $10^{10}$ & $10^{12}$ & $1.3$ & $6.5$ & $8.4\cdot10^4$ & $8.3\cdot10^4$\\
\hline
P${}_{13}$ & {\boldmath$10^{-8}$} & $0.1$ & $0.2$ & $10^{10}$ & $10^{12}$ & $23$ & $120$ & $2.6\cdot10^5$ & $2.5\cdot10^5$\\
\hline
P${}_{14}$ & {\boldmath$10^{-9}$} & $0.1$ & $0.2$ & $10^{10}$ & $10^{12}$ & $4.2\cdot10^2$ & $2.1\cdot10^3$ & $7.9\cdot10^5$ & $5.8\cdot10^5$\\
\hline
\end{tabular}
\caption{Number densities of warm axions calculated numerically for several benchmark points obtained using different approximations: without geometric and thermal corrections ($n_{CW}$), with only geometric corrections ($n_{CW+G}$), with only thermal corrections ($n_{CW+T}$) and with both types of corrections ($n_{CW+T+G}$). All densities are rescaled to one common temperature $T$ and are expressed in units of $T^3$. The results are presented with two significant digits. In some cases the third digit is added to indicate differences among results which are very similar. Points P${}_{8}$, P${}_{9}$ and P${}_{11}$ represent scenario B, point P${}_{7}$ represents scenario C1, all other points are of type C2 (see subsection \ref{sec:evolution_thermal} for definitions). 
The table is divided into five blocks. In each black only one parameter is changing (values written in bold face) with other parameters fixed.}
\label{table-BP}
\end{center}
\end{table}

The strong dependence of $n_{CW+T}$ (and $n_{CW+T+G}$) on $m/\mu$ in scenario C2 may be to some extend explained by the following reasoning. From eqs.~\eqref{DeltaV_SUSY_0} and \eqref{mS^2_SUSY_0} it follows that the number of particles which may be produced due to a tachyonic instability in the CW potential scales with the parameters roughly as $\lambda^{-1/2}m\mu^2$. Such production takes place when the temperature is close to the critical one \eqref{Tc} which is proportional to $m$.
Thus, if this happens after the reheating, the rescaled number density of warm axions is roughly proportional to
\begin{equation}
n_{CW+T},\,n_{CW+T+G} \propto \lambda^{-1/2}\left(\frac{m}{\mu}\right)^{-2}T^3\,.
\label{n-with-T}
\end{equation}
The dependence on the remaining parameters: $\delta$ (see points P${}_5$, P${}_8$ and P${}_{9}$) and especially $\mu$ and $H_I$ (see points P${}_1\div$P${}_3$) is much weaker.

Comparing \eqref{n-without-T} with \eqref{n-with-T} one can see that the way the number density of produced warm axions depends on the model parameters changes very much when the thermal corrections are taken into account.
$n_{CW}$ and $n_{CW+G}$ depend mainly on $\delta$ and $H_I$ while $n_{CW+T}$ and $n_{CW+T+G}$ depend mainly on the ratio $m/\mu$. 
In all cases there is a dependence on $\lambda$ but with thermal corrections it is weaker than without them (see benchmark points P${}_{12}\div$P${}_{14}$). 
Our numerical calculations showed that in big parts of the parameter space the approximations presented in this subsection lead to reasonable estimates of the results. However, in some other parts of the parameter space, e.g.~in the transition regions between scenarios C1 and C2 (close to green lines in Fig.~\ref{Fig-AS/S0}) or in scenario B, numerical calculations are necessary to estimate the results.

As the above arguments and the results from table \ref{table-BP} show, thermal corrections may change the number of warm axions even by several orders of magnitude. This number may be much smaller (in region C1, e.g.~point P${}_7$ for which the used approximation gives 0) or much larger (deep in region C2 with small $m/\mu$ and $H_I$, e.g.~roughly 7 orders of magnitude for point P${}_1$) than it is predicted with thermal effects neglected. However, it is not possible to arbitrarily increase the number of warm axions by decreasing $m/\mu$. As we explain in the next subsection, for small enough $m/\mu$ the saxion oscillations thermalize before any particles are produced. The limiting value of $m/\mu$ depends on other parameters of the model. Some examples are shown in Fig.~\ref{Fig-AS/S0} as horizontal blue lines. Any substantial number of warm axions may be produced only for parameters which are below a green line and above a blue line.

The geometric corrections increase the number of warm axions when the thermal corrections are neglected ($n_{CW+G}>n_{CW}$ for all benchmark points). With the thermal corrections included the geometric corrections  increase this number in scenario B but decrease it in scenario C2  ($n_{CW+T+G}>n_{CW+T}$ for points P${}_{8}$, P${}_{9}$ and P${}_{11}$ while $n_{CW+T+G}<n_{CW+T}$ for all other points). The reasons for such behavior were explained in subsection \ref{sec:evolution_thermal}.

So far in this subsection we dealt only with warm axions. One should remember that cold axions are also produced via a conventional misalignment mechanism. An interesting feature of our model is that 
for much of the parameter space the amount of cold axions is determined by stochastic processes during inflation like in models with broken PQ symmetry, despite the fact that here the PQ symmetry is unbroken for some time after the end of inflation. This unusual behavior follows from the dynamics of the PQ field. The saxion field oscillates keeping the information about the initial PQ field phase even during the period when the PQ symmetry is unbroken.

The sum of the energy densities of warm and cold axions should be compared with the observed density of DM. If the amount of warm axions is too small (or there are no warm axions at all like in cases C1 and D) it is possible in many cases to complement it with cold axions by choosing an appropriate phase of the PQ field generated during inflation. Thus, this model has extra flexibility in producing the correct amount of axion DM.


\subsection{Thermalization and production of other particles}
\label{sec:thermalization}

So far, when discussing parametric resonance, we concentrated on possible production of axions and saxions. But in principle other particles which couple to the oscillating saxion field may also be produced. In our model these are scalars $\phi_i$ and fermions $\psi_j$.
Fermions are not effectively produced due to their statistics -- strong resonant production may occur only for bosons. So, we should consider possibility of production of scalars $\phi_i$, especially during oscillations dominated by a quadratic thermal contribution to the saxion potential. A parametric resonance takes place if an appropriate 
adiabaticity condition is violated. In the considered case this may be written as an upper bound on the mass parameter $m^2$. The leading term of this bound reads
\begin{equation}
m^2 \lesssim \frac{\lambda}{6}\,\sqrt{n_{\rm eff}}\,TA_S\,.
\label{m2<}
\end{equation}
Of course, both the temperature, $T$, and the amplitude of saxion field oscillations, $A_S$, decrease with time. So, resonant production of scalars $\phi$ is not possible starting from the moment when the above inequality is violated for the first time.
The maximal possible value of the r.h.s.~of \eqref{m2<} is obtained by replacing $A_S$ with the initial amplitude of the order of $S_{\rm min,inf}$ \eqref{Smin^2_I} and replacing $T$ with $T_i$ corresponding to the Hubble parameter $H_i^{(T)}$ \eqref{H_i^T}. For example, if $T_{RH} > \widetilde{T}_{RH}$ this gives 
\begin{equation}
m^2 \lesssim \sqrt{6\lambda}\,H_i^{(T)}\,S_{\rm min,inf}\,.
\end{equation}
No scalars $\phi$ may be at all produced via a parametric resonance if the above condition is violated.
If it is fulfilled, numerical calculations are necessary to estimate the number density of produced scalars. In such a case one should check whether those produced scalars $\phi$ contribute too much to the total energy density of the universe. This should not be a problem in models in which scalars $\phi$ have non-negligible contribution to $n_{\rm eff}$. They are massive and are in (at least partial) contact with thermal plasma so their density will be much decreased due to the Boltzmann factor.

In some of the cases discussed in previous subsections saxions produced via a parametric resonance and/or oscillations of the saxion field surviving till late times may be unacceptable phenomenologically. One way to get rid of such unwanted relics is their thermalization. Thermalization of oscillating scalar fields was analyzed in \cite{Mukaida:2012qn}. Due to interactions with thermal plasma the equation of motion \eqref{eomS}
is modified to 
\begin{equation}
\ddot{S}+\left(3H+\Gamma_{\rm th}\right)\dot{S}+\frac{\partial V_{\rm eff}}{\partial S}=0\,,
\label{eomST}
\end{equation}
where the dissipative coefficient has contributions from interactions with scalars and with fermions: $\Gamma_{\rm th}=\Gamma_{\rm th}^{(\phi)}+\Gamma_{\rm th}^{(\psi)}$. These coefficients depend on temperature in different way \cite{Mukaida:2012qn}:\footnote{For big enough amplitude of oscillations the fermion contribution may be dominated by a term of the form characteristic for scalars, $\lambda^2S^2/T$. However, as we will show later, such term is much less important for the thermalization process in our model.}
\begin{equation}
\Gamma_{\rm th}^{(\phi)}\sim \frac{\lambda^2 S^2}{\alpha_{\rm th}T}\,,
\qquad\qquad
\Gamma_{\rm th}^{(\psi)}\sim y^2\alpha_{\rm th}T\,,
\label{Gamma_th}
\end{equation}
where $\alpha_{\rm th}$ is an effective coupling of $\phi$ or $\psi$ to the thermal bath. In the case of the QCD axion this coupling is of order of the strong coupling $\alpha_s$ (renormalized at an appropriate energy scale) while for a general ALP it may be much smaller and is model dependent.

The scalar term $\Gamma_{\rm th}^{(\phi)}$ is proportional to the square of the saxion field at a given time.
One could expect that this term is more efficient for thermalization of saxion oscillations when the amplitude of these oscillations is large. In addition, the explicit temperature dependence enhances this contribution at later times. However, such expectations are not correct. One should observe that $\Gamma_{\rm th}^{(\phi)}$ is not proportional to the square of the amplitude of saxion oscillations but to a momentary field value. For $\Gamma_{\rm th}=\Gamma_{\rm th}^{(\phi)}$ the thermal dissipative term in \eqref{eomST} is proportional to $S^2\dot{S}$ so it vanishes both at maximal and minimal $S$. We checked by numerical simulations that this term does not lead to quick evaporation of oscillations. It only makes the amplitude to decrease faster than in the case with only Hubble friction $3H\dot{S}$ in \eqref{eomST}. On the other hand, the fermion contribution \eqref{Gamma_th} does lead to evaporation of oscillations when it becomes bigger than the Hubble parameter.

The saxion thermal mass is proportional to $m_S^T\sim\sqrt{\lambda n_{\rm eff}}\,T$ while for our quasi-supersymmetric spectrum $\Gamma_{\rm th}^{(\psi)}\sim \lambda\alpha_{\rm th}T$. The ratio ${\Gamma_{\rm th}^{(\psi)}}/{m_S^T}$ is proportional to $\sqrt{\lambda}$ so typically it is very small. Thus, thermalization may take place only long after the onset of saxion oscillations because the Hubble parameter must decrease by a factor of order $\sqrt{\lambda}\ll1$. This happens at temperature of order 
\begin{equation}
T_{\rm th} \sim \sqrt{\frac{45}{4\pi^3g_*}}\,\alpha_{\rm th}\lambda{M_{Pl}}\,.
\end{equation}
It is important for the phenomenology of a given model whether thermalization occurs after or during processes described in section \ref{sec:evolution_thermal}. In the latter case those processes terminate at the moment of thermalization when still ongoing oscillations evaporate. This of course changes the later evolution. 
So, we have to consider the forth scenario in addition to the three discussed in section \ref{sec:evolution_thermal}.

\begin{itemize}
\item[{\bf D:}] {\bf Early thermalization: {\boldmath $T_{\rm th}>\widetilde{T}$}}

Thermalization of saxion oscillations happens during time when the  potential of the PQ field is still dominated by the thermal mass term if $T_{\rm th}$ is bigger than $\widetilde{T}$ given by \eqref{Ttilde}. 
This condition may be written in the following form
\begin{equation}
\frac{m^2}{\mu^2}\,\ln\left(\frac{e\mu^2}{m^2}\right)
\lesssim
\frac{15n_{\rm eff}\big(\widetilde{T}\big)}{\pi g_*N_s}\,
\alpha_{\rm th}^2\lambda^2\frac{M_{Pl}^2}{\mu^2}\,.
\label{Tth>Ttilde}
\end{equation}
When it is fulfilled the saxion oscillations evaporate without producing any substantial number of saxions and axions because saxion field oscillated only in approximately quadratic potential. Moreover, after thermalization the PQ field vanishes so later the model behaves as the standard ``unbroken ${U(1)_{PQ}}$'' one. The PQ symmetry is broken at some lower temperature what leads to the white noise power spectrum of the axion field.

The expression on the l.h.s.~of \eqref{Tth>Ttilde} goes to zero when $m^2/\mu^2\to0$ or $m^2/\mu^2\to e$. So, there are always values of $m^2/\mu^2$ for which the early thermalization takes place. If the r.h.s.~of \eqref{Tth>Ttilde} is bigger than 1 then condition \eqref{Tth>Ttilde} is fulfilled for any $m^2/\mu^2$. 
As an example let us consider a QCD axion for which the coupling $\alpha_{\rm th}$ is somewhat below 0.1 (SU(3) coupling at high energy scale) and $n_{\rm eff}$ may be close to $\frac32N_s$.  
In such a case the r.h.s.~of \eqref{Tth>Ttilde} is bigger than 1 if $\mu\lesssim\lambda\cdot 10^{17}\,$GeV. In models with small enough $\mu$ early thermalization of QCD saxion oscillations is realized for any value of $m$.
\end{itemize}
The blue horizontal lines in Fig.~\ref{Fig-AS/S0} are related to the above described scenario D. Early thermalization of oscillations takes place for parameters below a given blue line if the r.h.s.~of \eqref{Tth>Ttilde} has value equal to the label of that line\footnote{Combination $(m^2/\mu^2)\ln(e\mu^2/m^2)$ takes a given small value not only for small $m^2/\mu^2$ but also when $m^2/\mu^2$ is slightly smaller then $e$. The corresponding blue lines are not shown in Fig.~\ref{Fig-AS/S0} because they would be indistinguishable from the dotted lines. Scenario D may in principle be realized above such blue lines but this requires strong tuning of the ratio $m^2/\mu^2$.}. 
When $T_{\rm th}<\widetilde{T}$, i.e.~when condition \eqref{Tth>Ttilde} is violated, the PQ field evolves according to one of the scenarios -- A, B or C -- described in section \ref{sec:evolution_thermal} and only later energy stored in saxion particles and oscillations is thermalized. In all cases thermalized saxion may contribute to adiabatic perturbations but not to isocurvature ones.

Sometimes a Higgs portal coupling is introduced to thermalize some unwanted scalar oscillations \cite{Co:2020dya}. However, as we argue below eq.~\eqref{Gamma_th}, this is not an efficient way to do so (for oscillations in the potential with the minimum located at vanishing or very small field value). In our model we do not need to add any new ingredients to the model to achieve thermalization. The PQ field does couple to some fermions. There are models (e.g.~those with the QCD axion) in which at least some of those fermions may be in contact with thermal plasma.
Such fermions not only contribute to the thermal mass of the PQ field (what may be crucial for the evolution of saxion oscillations and particle production) but may also lead to thermalization of saxion oscillations. From this point of view our model is more economical.


\section{Conclusions}
\label{sec:Conclusions}

In this paper we have investigated models with a Peccei-Quinn-like field, $\Phi$, which has a very small self-coupling. We found that dynamics of such field during and after inflation is quite non-trivial. A very important role is played by different kinds of corrections to the tree-level potential: radiative, thermal and those caused by the curvature of space-time during early stages of the evolution of the universe. With radiative corrections taken into account it is very natural that the Peccei-Quinn ${U(1)_{PQ}}$ symmetry is broken by the Coleman-Weinberg mechanism instead of usually  assumed tree-level ``Mexican hat'' potential.

We checked that corrections related to the space-time geometry may very strongly change the  character of the PQ field evolution during inflation. $\Phi$ is a light scalar spectator field so it undergoes quantum fluctuations in the nearly de Sitter inflationary space-time. In many models, in order to fulfill constraints on the isocurvature perturbations, those fluctuations must cumulate in order to make $\Phi$ very large in the observed part of the universe. Usually this is achieved by assuming very long inflation and extremely small $\Phi$ self-coupling. Situation is quite different in our model with the CW potential with geometric correction taken into account. If the Hubble parameter during inflation is not too small the resulting potential develops a new relatively deep minimum at some large value of $\Phi$. 
The PQ field at the end of inflation has large value due to the classical motion rather than due to large quantum fluctuations. Especially interesting in this regard are models in which particles to which the PQ field couples have a quasi-supersymmetric spectrum.

Evolution of the PQ field after inflation may be quite complicated and quite different for different regions of the parameter space. For some time after the end of inflation, due to the Hubble friction, both components of complex $\Phi$, i.e.~the radial saxion field $S$ and the angular axion field $a$, stay frozen with values generated stochastically during inflation. At some point the saxion field, which is massive, starts to oscillate. We checked that typically (for not very large values of $H_I$) 
these oscillations start due to the thermal correction. In models with a tree-level potential saxion and axion particles are quite efficiently produced during such oscillations by the parametric resonance mechanism. 
Situation may be very different in models investigated in this paper. With thermal corrections taken into account quite often the resonant production of saxions and axions is very strongly suppressed. Such particles may be later produced due to a different mechanism, namely due to the tachyonic instability. This conclusion applies also to models with a simple tree-level PQ potential. The number density of warm axions produced this way depends strongly on some parameters of our model. It may be even orders of magnitude smaller or bigger than in the case with thermal corrections neglected. Especially many warm axions are produced if the full potential with the corrections has for some range of temperatures a barrier separating the global minimum from a local one. 
We discussed all these properties of the model using approximate analytical expressions and presented also results of numerical calculations for several benchmark points. 
Those results for the production of axions have phenomenological consequences because axions produced this way may contribute to the relic density of warm dark matter or dark radiation.

The axion field in our model contributes also to the relic density of cold dark matter due to the standard misalignment mechanism. There are two possibilities which may be realized in different parts of the parameter space. For some values of the parameters the relic density of axion CDM depends only on the axion decay constant and does not depend on the axion field value just after inflation. The power spectrum of the axion CDM density has characteristic of white noise at relatively small scales. These are features typical for ``unbroken PQ symmetry'' models.
For other values of the parameters the relic density of axion CDM does depend on the initial value of the axion field just after inflation.
This and also the power spectrum are typical for ``broken PQ symmetry'' models.

Finally we considered thermalization of possible residual saxion oscillations and produced saxion particles. We argued that thermalization due to couplings of the PQ field to scalars, also to the Higgs scalar, is not efficient. On the other hand, thermalization, sometimes necessary from the phenomenological point of view, may be realized via couplings to fermions. Generically the PQ fields couple to some fermions so no extra ingredients are necessary to have thermalization in our model, of course if at least some of such fermions have sufficient couplings to some particles in the thermal bath (as e.g.~in models of the QCD axion).

Dynamics of saxion and axion fields in our model with different kinds of corrections taken into account may be very rich and interesting. It may have some features of the ``unbroken PQ symmetry'' as well as ``broken PQ symmetry'' models. However, this dynamics is rather complicated and dedicated numerical calculations are usually necessary to obtain quantitative predictions.

\section*{Acknowledgments}
Work partially supported by National Science Centre,
Poland, grant DEC-2018/31/B/ST2/02283.



\bibliographystyle{BiblioStyle}
\bibliography{axions_JCAP2}

\end{document}